%% file: 0_main.tex
\documentclass[letterpaper,twocolumn,10pt]{article}
\usepackage{usenix-2020-09}

\AtBeginDocument{%
  \providecommand\BibTeX{{%
    \normalfont B\kern-0.5em{\scshape i\kern-0.25em b}\kern-0.8em\TeX}}}

\usepackage{tikz, amsmath, filecontents, algorithm, amsmath,epsfig,endnotes, enumitem,makecell,multirow, microtype, xspace, setspace, xcolor,textcomp,graphicx,filecontents,caption,subcaption, url,tabularx, graphicx, times}
\usepackage[noend]{algpseudocode}
\usepackage[compact]{titlesec}

\begin{document}

\newcommand{\remove}[1]{}
\newcommand{\staticgraph}[0]{\emph{DFG}\xspace }
\newcommand{\dynamicgraph}[0]{\emph{ADFG}\xspace }
\newcommand{\schedulequeue}[0]{\emph{scheduling queue}\xspace }
\newcommand{\execqueue}[0]{\emph{execution queue}\xspace }
\newcommand{\statesMonitor}[0]{Global State Monitor\xspace }
\newcommand{\CompassCache}[0]{\emph{Compass cache}\xspace }

\newcommand{\ay}[1]{\noindent{\textcolor{magenta}{\bf \fbox{AY} {\it#1}}}}
\newcommand{\andrea}[1]{\noindent{\textcolor{brown}{\bf \fbox{AM} {\it#1}}}}
\newcommand{\rv}[1]{\noindent{\textcolor{red}{\bf \fbox{RV} {\it#1}}}}
\newcommand{\tc}[1]{\noindent{\textcolor{blue}{\bf \fbox{TC} {\it#1}}}}

\title{\Large \bf Compass: A Decentralized Scheduler for \\ Latency-Sensitive ML Workflows}

\author{
{\rm Yuting Yang}\\
Cornell University\\
USA\\
yy354@cornell.edu
\and
{\rm Andrea Merlina}\\
University of Oslo\\
Norway\\
andremer@ifi.uio.no
\and
{\rm Weijia Song}\\
Cornell University\\
USA\\
ws393@cornell.edu\\
\and
{\rm Tiancheng Yuan}\\
Cornell University\\
USA\\
ty373@cornell.edu
\and
{\rm Ken Birman}\\
Cornell University\\
USA\\
ken@cs.cornell.edu
\and
{\rm Roman Vitenberg}\\
University of Oslo\\
Norway\\
romanvi@ifi.uio.no
}







\maketitle

\begin{abstract}


We consider ML query processing in distributed systems where GPU-enabled workers coordinate to execute complex queries: a computing style often seen in applications that interact with users in support of image processing and natural language processing.  In such systems, coscheduling of GPU memory management and task placement represents a promising opportunity.  We propose Compass, a novel framework that unifies these functions to reduce job latency while using resources efficiently, placing tasks where data dependencies will be satisfied, collocating tasks from the same job (when this will not overload the host or its GPU), and efficiently managing GPU memory.  Comparison with other state of the art schedulers shows a significant reduction in completion times while requiring the same amount or even fewer resources.  In one case, just half the servers were needed for processing the same workload.

\end{abstract}

\input{1_introduction}

\input{2_background}

\input{3_system_architecture}

\input{4_compass}
\input{5_implementation}
\input{6_experiments}
\input{7_related_work}
\input{8_conclusion}
\input{10_Acknowledgments}

\bibliographystyle{plain}

\input{bibliography}

\end{document}

%% file: 1_introduction.tex
\section{Introduction}
\label{sec:introduction}


The work described here is motivated by a recent trend:  under pressure to respond rapidly to incoming events, the cloud is expanding to include edge clusters physically close to the end-user.  Our new scheduler, Compass, targets this setting, managing resources in ways that improve job completion delay with the same (or reduced) level of resources.  Machine intelligence is already popular in interactive applications, and projected to gain importance for at least another decade, so progress could be broadly valuable.  

Intelligent interactive services are not real-time systems in the classic sense.  In fact, they more closely resemble cloud microservice architectures: a single input event will often trigger a pipelined computation in which multiple ML models play distinct roles.   
The end user wants quick responses hence the hosting infrastructure must minimize avoidable overheads: a goal shared with the first tiers of the cloud.  Yet intelligent edge applications differ from cloud microservices in important ways, so we cannot just use the same techniques employed in web frameworks. Whereas the outer tiers of today's cloud are dominated by lightweight, stateless, containerized applications that can be upscaled or downscaled at low cost, ML depends on large objects (hyperparameters, model parameters, and supporting databases) and often entails hardware-accelerated computation using devices preconfigured with the proper firmware.  When shifting a task to a device that has not previously run it,  computation cannot begin until all the prerequisites are in place.   We can and do launch new ML instances when additional capacity is needed, but scheduling strategies must evolve to avoid thrashing.


Data dependencies also pose a challenge.  Host memories are large but GPU memories are smaller and expensive.  ML model parameters ("ML models" for short) can be hundreds of megabytes in size~\cite{megatron}.  Substantial delays arise if a required model isn't resident in GPU memory when an ML task is launched.  This leads us to treat GPU memory as a cache and to view the cache hit rate as an important metric.

Our work represents edge ML applications as acyclic data-flow graphs ({\staticgraph}s). Each ML instance runs in its own address space hence a single application could include ML pipelines implemented using different tools  (we currently support PyTorch~\cite{torch}, TensorFlow~\cite{tensorflow}, and MXNet~\cite{MXNet}).  A triggering event launches one job, and each {\staticgraph} step within it runs as a task on a single server (we will relax these restrictions in future work).  Compass plays two roles: platform-level GPU cache management and job/task placement. Our algorithm has the primary goal of minimizing job latency, with secondary goals of achieving efficient hardware utilization and packing work into as few servers as feasible so that unused hardware can be put into a power-saving mode. 

An important decision was to make  Compass fully decentralized and symmetric: every worker node can schedule tasks on every other worker. While many existing schedulers~\cite{centralizedCoreGranularity, InferLine, Clockwork, antman, singularity, unearth_inter_job_dependency} are centralized, permitting fully coordinated decision-making, centralization introduces overheads due to the need to consult the scheduler and to notify it of changing conditions, creating possible bottlenecks.  Early in our effort, we realized that in edge systems task execution times are not fully predictable, hence worker loads and GPU memory contents can quickly evolve in unanticipated ways. Decentralized tracking of system state was traditionally assumed to be far more costly, but modern data replication is inexpensive and scalable~\cite{Derecho}.  Our decentralized solution has low overheads and outperforms centralized alternatives.

The main contributions of this paper include:

\setlist{nolistsep}
\begin{enumerate}[noitemsep]
    \item A decentralized scheduler that reduces end-to-end latency for ML applications on edge clusters by anticipating which ML models will be needed by each GPU. 
    
    \item GPU memory management that leverages {\staticgraph}s to optimize GPU cache hit rates.
    
    \item A decentralized service for tracking server loads and GPU cache contents that imposes negligible overheads yet enables significant scheduling improvements.  
    
    \item Experiments and simulation showing that  Compass can reduce latency by 2x-6x.  Compass never consumes more resources than existing schedulers and requires as few as half the servers to run an identical workload.
    
\end{enumerate}

%% file: 2_background.tex
\section{Deployment Scenarios and Environment}

\subsection{Dataflow Graphs} 
\label{sec:DataflowGraph}
\begin{figure}[ht]
     \centering
     \begin{subfigure}[b]{0.45\textwidth}
         \centering
         \includegraphics[width=\textwidth]{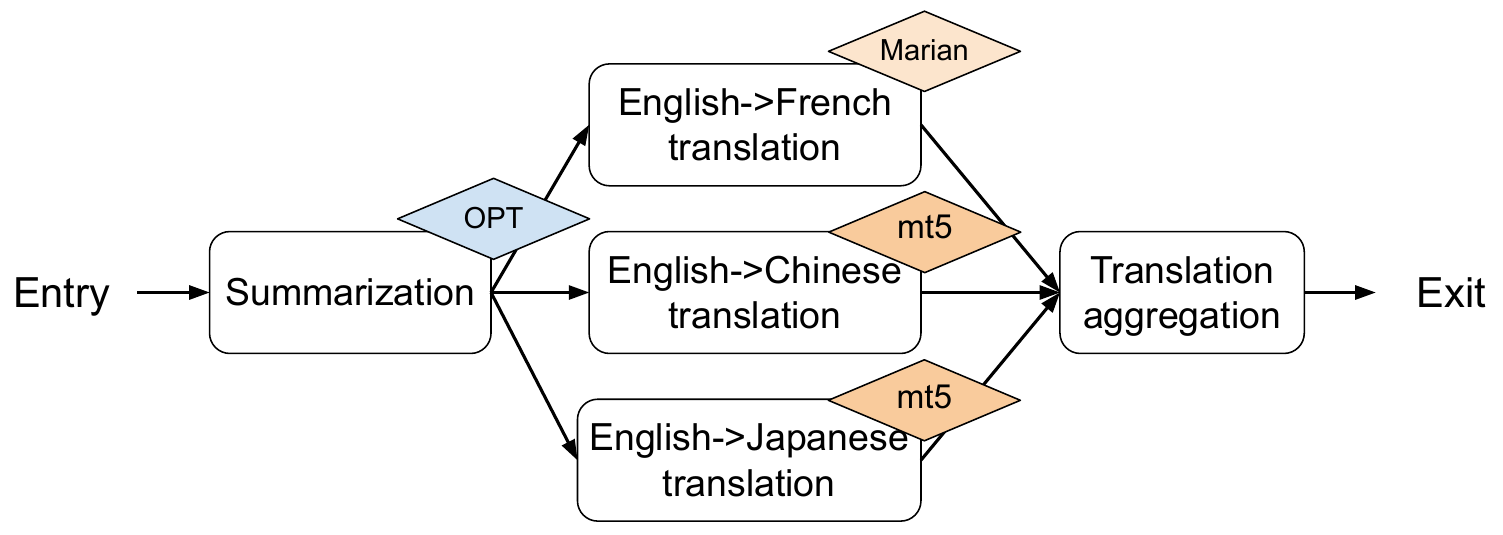}
         \caption{language translation pipeline}
         \label{fig:translation}
     \end{subfigure}
     \hfill
     \begin{subfigure}[b]{0.45\textwidth}
         \centering
        \includegraphics[width=\textwidth]{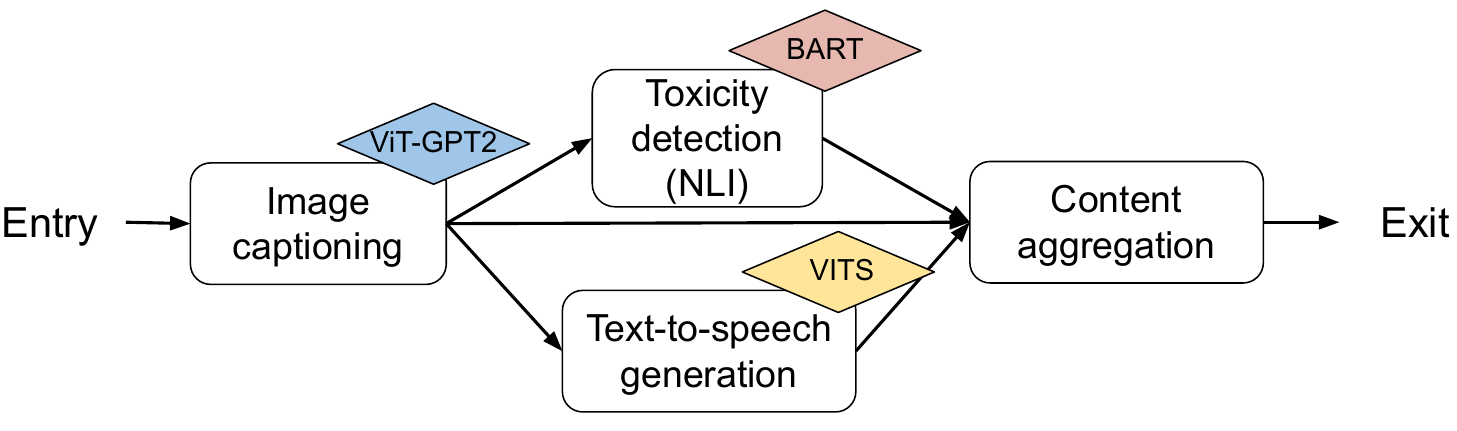}
         \caption{image reading pipeline}
         \label{fig:readimg}
     \end{subfigure}
     \hfill
     \begin{subfigure}[b]{0.23\textwidth}
         \centering
         \includegraphics[width=\textwidth]{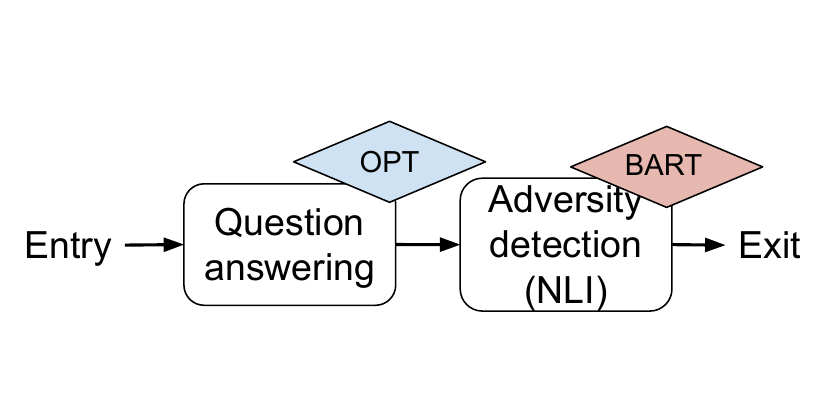}
         \caption{Q\&A dialogue pipeline}
         \label{fig:dialogue}
     \end{subfigure}
     \hfill
     \begin{subfigure}[b]{0.24\textwidth}
         \centering
         \includegraphics[width=\textwidth]{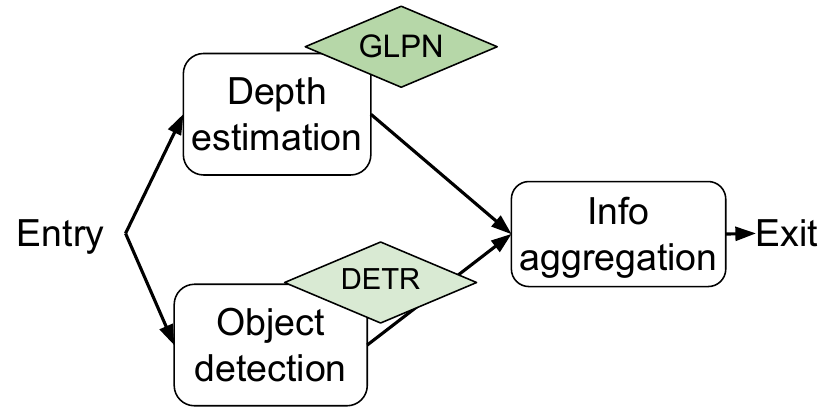}
         \caption{3D perception pipeline}
         \label{fig:3dperception}
     \end{subfigure}
\caption{Pipelines}
\label{fig:pipelines}
\end{figure}

Compass's \staticgraph representation is similar to that used in prior work~\cite{Quincy, Apollo, orion, InferLine}. As noted earlier, these are directed acyclic graphs $G=(E,V)$, where vertex $v \in V$ represents an ML computation.  Edge $e \in E$ represents precedence constraints:  output from the upstream ML will become input to the downstream ML. We attach a diamond box to vertex $v$ to represent {\em data dependencies}: the ML model and other objects that $v$ requires to perform its computational task. As seen in Figure~\ref{fig:pipelines}, identical colors denote the same ML model, while different ones denote different models. Our figures omit details such as object sizes and estimated job/task execution times, but the \staticgraph also includes them. 
Our experimental section used these workflows as its scheduling target.

The workflow in Figure~\ref{fig:translation} autocaptions for multilingual online meetings~\cite{multilanguageOCR}, aggregating the translations as a single output.  It uses Meta's opt-1.3b Open Pre-trained Transformer (OPT) model to process input~\cite{opt}, Marian~\cite{marian, en_fr_model} for French translation, and mt5~\cite{mt5, zh_ja_en_model} for both Chinese and Japanese (mt5 plays two roles but uses a single model).

The second workflow (Figure~\ref{fig:readimg}) arises in an application that auto-captions images for children's education~\cite{education}. The {\staticgraph} employs the ViT-GPT2 model~\cite{vit-gpt2, vit-gpt2-model} for automated captioning,  ESPnet~\cite{espnet} for vocalization, and BART~\cite{bart} to ensure that only child-safe results are generated. 

The third workflow (Figure~\ref{fig:dialogue}) might be 
 used by a Virtual Personal Assistant (VPA)~\cite{personal_assistant}; it uses the same OPT model but with prompts to "shape" the desired output, configuring BART to target an adult  rather than a child.

The fourth workflow (Figure~\ref{fig:3dperception}) is intended to assist a vision-impaired user.  Object detection is performed by the DEtection TRansformer (DETR)~\cite{detr} model and depth estimation using a hierarchical transformer encoder-decoder model~\cite{glpn}. The final vertex combines these two estimates.

\subsection{Deployment Assumptions} 
\label{sec:Deployment Assumption}
Compass could be used in a cloud datacenter, but will often be deployed in edge clusters located near data-capture devices, an approach also adopted in prior work ~\cite{ekya}. These clusters would be more limited than full-fledged cloud servers but still could host multiple {\staticgraph}s to serve mixes of edge requests.  Using secure cloud links, edge applications would handle time-sensitive tasks while drawing on the full cloud for aspects off the low-latency critical path.

A single GPU memory would often be too small to hold the full set of active ML models.  The MLs (graph vertices) in Figure~\ref{fig:pipelines} depend on objects that are each several GB in size.  The total memory aggregated over the full set of {\staticgraph}s is nearly 35GB, which already exceeds what a single standard cloud GPU could hold\footnote{We additionally assume that the GPUs have been preconfigured with all required computational libraries, but these need to remain resident in GPU memory and hence are not a target for optimization.}.  Of course, one could find GPUs with more memory, but one could also find mixes of workflows that depend on more ML models (moreover, GPU memory is costly and it is unlikely that edge systems would have the most expensive GPU devices).  In contrast, host memory is often far larger.  For example, Intel's Xeon Gold 6242R Processor can be configured with 1TB of DRAM~\cite{xeon_gold_6242}, and the  E5-2403 normally ships with 384GB DRAM~\cite{intel_e5}.  Thus GPU memory is the potential bottleneck.   

The {\staticgraph}s describing workflows that the system might encounter in a particular deployment are small and static, hence we assume that these are available on all workers. To annotate our  {\staticgraph}s with data about expected runtimes and model sizes, we profiled representative test cases, then adopted parameters covering 95\% or more of the observed data.


\subsection{Scheduler Objectives} 
\label{sec:objectives}
Many ML schedulers~\cite{themis, ekya, pollux, heterogeneityAwareClusterSchedule} focus on long-running distributed training tasks, but edge systems are dominated by event classification rather than training, and would not typically include this style of iterative computation. Distributed patterns such as AllReduce still arise, but we treat these parallel tasks exactly like other kinds of parallel tasks.

Our four example ML workflows all require fast responses. For example, when performing language translation and image generation, end-to-end performance of the {\staticgraph} determines the wait time the human user will perceive.  
Compass's primary role is to minimize end-to-end latency, defined as the difference between the end time (when the processing of the exit task is finished) and the start time (when the job instance is generated). 
Minimizing latency in \staticgraph structured workflows is NP-complete, hence Compass will inevitably depend on heuristics~\cite{nptheory, dag_scheduling, npcomplete}. 
We do not explicitly optimize for resource consumption or energy, but our algorithm turns out to be parsimonious in both respects.

%% file: 3_system_architecture.tex
\section{System Architecture} 
\label{sec:System Architecture}
\begin{figure}[b]
    \centering
    \includegraphics[width=0.9\columnwidth]{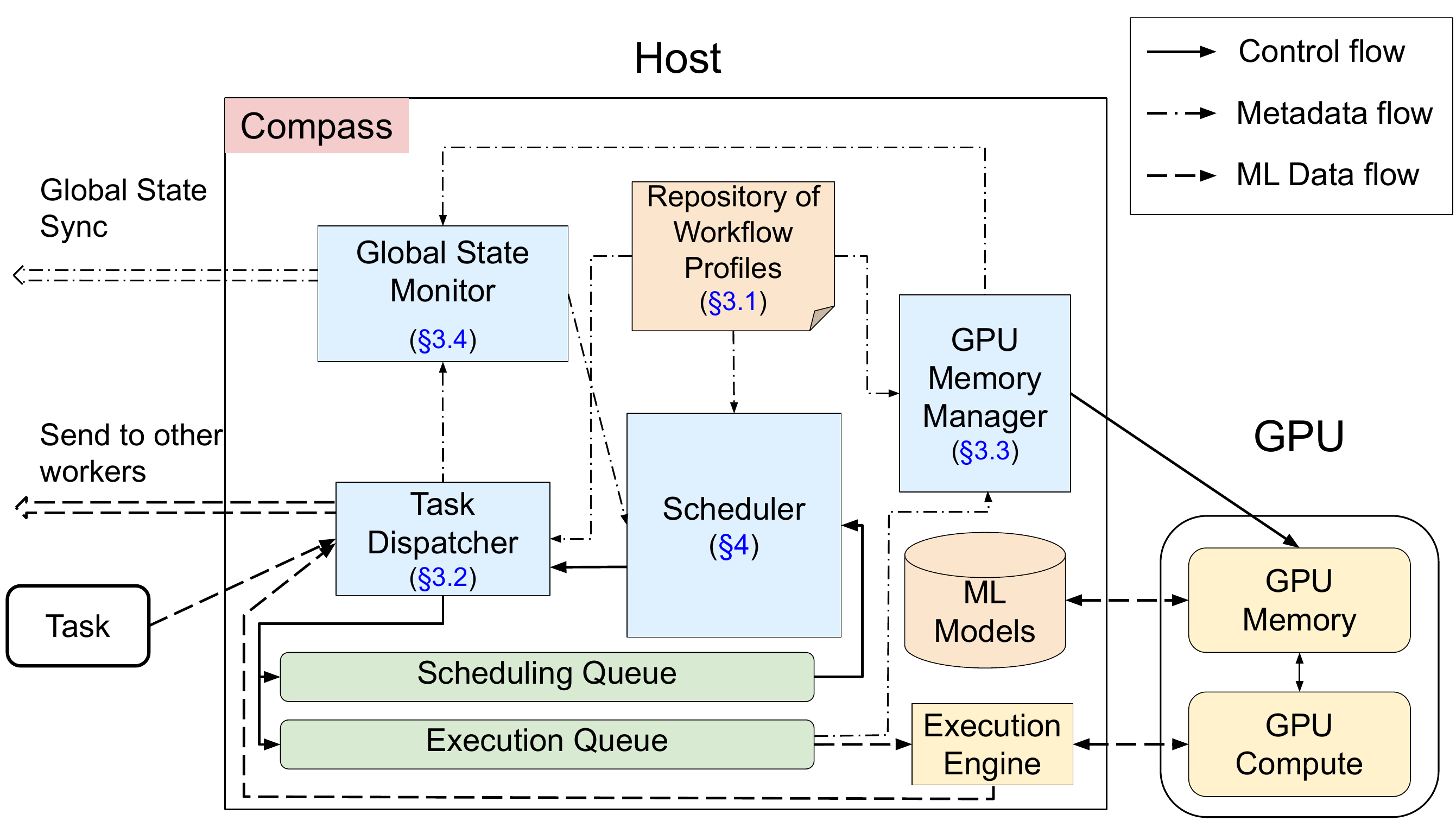}
    \caption{Worker Components in Compass}
    \label{fig:system architecture}
\end{figure}

As shown in Figure~\ref{fig:system architecture}, we assume that each worker consists of a host supporting storage, host computation, and a GPU unit for heavy lifting. All workers run Compass, which is a fully decentralized and symmetric: every worker has visibility of the system state and can assign tasks to any of its peers. 

Job instances are scheduled and executed by workers at the granularity of {\em tasks}. A client triggers a new job by sending a request containing an input object to one of the workers, which launches the ingress task. As its first step, this task creates an \emph{Activated Dataflow Graph} (\dynamicgraph) for the job instance. \dynamicgraph is based on the \staticgraph; it is a map containing initial worker assignment for all tasks in the job instance. However, the assignment can still be dynamically adjusted, as we explain below. Once created, the \dynamicgraph is piggybacked from task to task as the job executes.

The architecture of Compass is seen in Figure~\ref{fig:system architecture} and consists of the following components: Workflow Profiling collects profiled information about the vertices in a \staticgraph such as average execution times. The Decentralized \statesMonitor collects and periodically exchanges information about all the workers: their load, cache bitmap, etc. Compass's Scheduler component creates a \dynamicgraph and later adjusts it. The task dispatcher executes the decisions taken by the Scheduler, sending task start requests to the correct worker as the prior task(s) complete.  The GPU Memory Manager fetches models on the fly and manages a GPU cache. Task processing is managed by the Execution Engine, which includes a plug-in customized for each supported ML framework. We now describe each of the components, while the details of Compass's scheduling algorithm are presented in Section~\ref{sec:CompassSchedulerDesign}.

\subsection{Repository of Workflow Profiles} 
\label{sec:Workflow Profiles}
Compass's Workflow Profiles Repository holds meta-information about  {\staticgraph}s. 
In addition to static information (expected runtime costs and input/output object sizes), each task is annotated with the  actual sizes of known inputs. 

\subsection{Scheduler and Task Dispatcher} 
\label{sec:Task Dispatcher}

\begin{figure}[b]
    \centering
    \includegraphics[width=0.85\columnwidth]{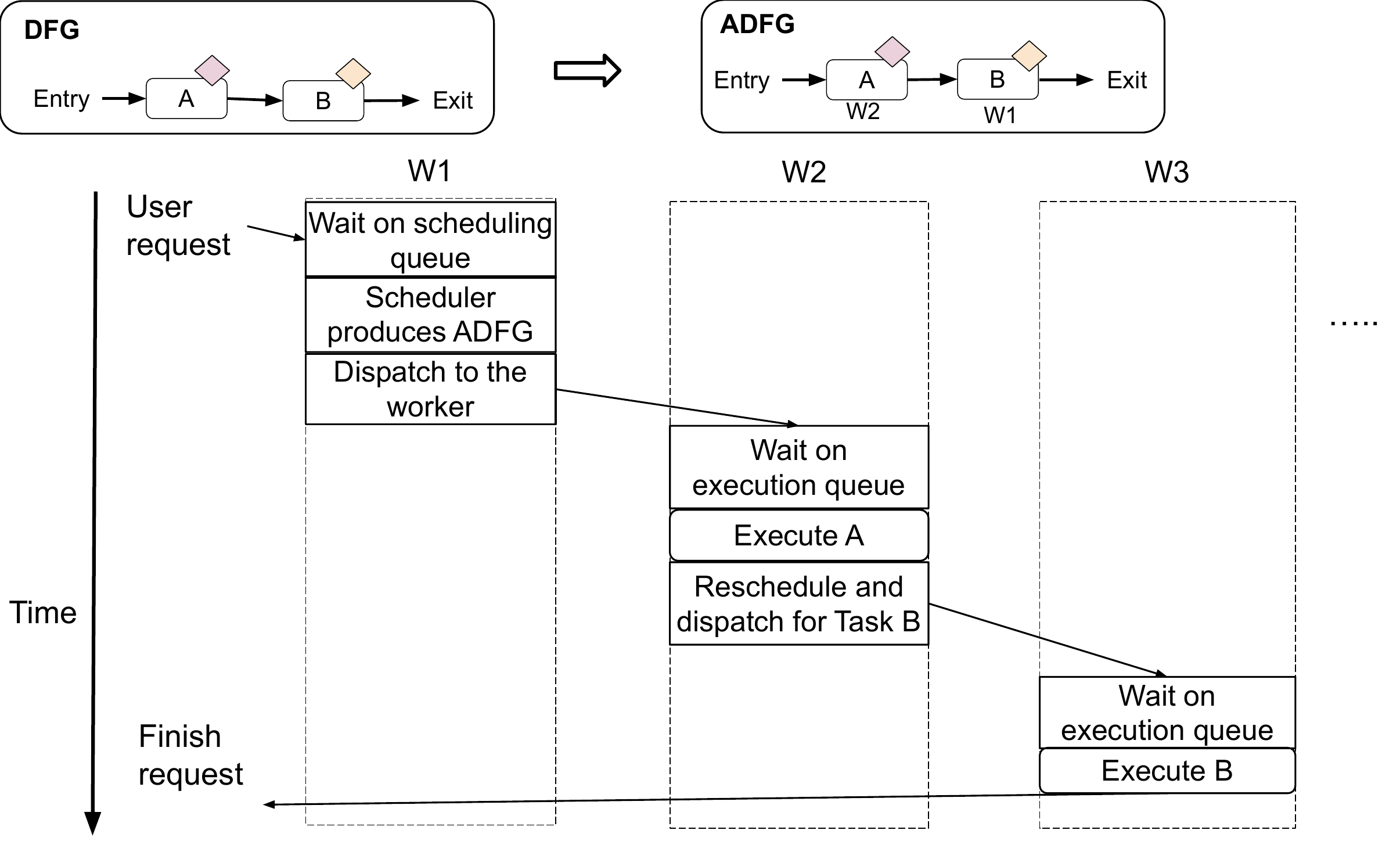}
    \caption{Example of Job Instance Handling}
    \label{fig:DispatchScheduleOverview}
\end{figure}


The flow of job handling is illustrated in Figure~\ref{fig:DispatchScheduleOverview}. When a worker $w$ receives a job processing request from a client, the request is appended to \schedulequeue on $w$. The Scheduler component loops, processing the first request on the \schedulequeue. For each request, it produces an \dynamicgraph which includes the worker assignment for each task.

Once the \dynamicgraph has been created, Task Dispatcher on $W$ sends it to worker $v$ on which the entry task in \dynamicgraph is scheduled for execution ($v$ may happen to be $w$). Upon receiving the \dynamicgraph, $v$ appends the entry task to its \execqueue. The Task Dispatcher on $v$ also loops, examining the first task $t$ on the \execqueue, popping it from the queue if it can be executed, and then  repeating as soon as the worker has adequate resources to process a new task. 

The determination of readiness for execution  entails verifying that all inputs to $t$ produced by predecessor tasks in the \dynamicgraph are available. If not, Task Dispatcher temporarily leaves $t$ in the queue and proceeds to the next task. 


Once the prerequisites for executing $t$ are satisfied, the next step entails ensuring that the ML model is in GPU memory. If the model is not in the GPU cache, it will be transferred in from host memory. While the transfer is underway, Task Dispatcher leaves $t$ in the queue and proceeds to the next task.

Once the model is in the GPU cache, the Execution Engine starts processing $t$ by doing an upcall to the task logic in the relevant framework. When the execution is finished, if necessary, the Scheduler adjusts the worker assignment for successors of $t$ in the \dynamicgraph using the algorithm presented in Section~\ref{sec:CompassSchedulerDesign}. After the adjustment,
Task Dispatcher sends the \dynamicgraph along with the output data of $t$ to workers assigned to execute the successors of $t$.

Preliminary scheduling decisions enable Compass to collocate a series of tasks on the same worker.  This saves time by avoiding the need to move the output of one task to the location where the next worker would run. Moreover some aspects of task assignment can only occur at the outset:  if a set of tasks are followed by a join, they would have no way to make a coordinated assignment for the join task. But the unpredictability of actual runtimes and object sizes can create situations in which rigidly adhering to initial scheduling decisions could overload a worker.  Accordingly, we adopted a two-step scheduling approach in which the initial \dynamicgraph can be adjusted during runtime.  This turns out to be essential (see Section~\ref{sec:AblationFeatureAnalysis}).

Compass's Task Dispatcher also keeps track of the tasks it has accumulated on its local \execqueue, and estimates the wait time on this worker for a newly arriving task based on its current queuing situation. Task Dispatcher disseminates this information to all other workers through \statesMonitor. Schedulers on other servers could use this information to produce task assignments.

\subsection{GPU Memory Manager} 
\label{sec:GPUMemoryManager}
The purpose of this component is to manage ML models on worker nodes: the \CompassCache and the GPU model execution memory.  \CompassCache holds reusable model objects in compressed form, whereas the execution memory is used to hold input objects for tasks and a decompressed model for each currently-active task.


As mentioned in Sections~\ref{sec:introduction} and~\ref{sec:Deployment Assumption}, GPU memories will often be too small to simply hold copies of every ML model that might be needed, and yet it is costly to fetch large models at the last instant when a task will be executed.   
Fortunately, ML models are reusable across different instances of the same pipeline and even across different pipelines, as we saw in Figure~\ref{fig:pipelines} where a single model was able to produce output in multiple languages.
Retention of models in GPU memory can thus have a significant impact on job completion times. Prior work has often employed a static model placement, designating some set of workers that will "serve" each model, and managing GPU memory using a policy that the ML logic itself ultimately controls. In contrast, we take the approach of scheduler-triggered GPU memory management, in which the scheduler decides which worker will run each task, at which point the worker itself makes local decisions about model placement (both fetching and eviction) based on its assigned tasks. This design is especially beneficial for dynamic workloads with high variation or a sudden burst of requests for the same pipeline, because the Compass scheduler can dynamically add additional workers to accommodate bursts of demand for particular MLs. 

In the experiments in Section~\ref{sec:Experiments and Evaluation}, we can see that proper task assignment depends upon careful management of model placement, and that Compass is able to adjust to workloads with variations. 
While Compass's fetching and eviction policies are configurable, an important design element is to look ahead in the \execqueue and consider what policies will be required for the currently queued tasks. 
We number active ML models in a small id space (currently 0..63), enabling the
GPU Memory Manager to publish a bitmap encoding the current contents of the \CompassCache for each worker.  The effect is that all workers know the cache contents of all their peers and can take this into consideration when making task assignments, thereby opportunistically reusing models already in GPU memory.

\subsection{{\statesMonitor}} 
\label{sec:GlobalStateMonitor}
Compass's
\statesMonitor manages a distributed table of system state.  It runs on all workers and holds per-worker queue processing time and GPU memory (cache) contents. The queue processing time is the time to complete all tasks currently on the worker's \execqueue. The Compass cache information includes the bitmap specifying which models are in the cache, as well as the amount of free memory in the cache.

To distribute updates, we take advantage of a modern multicast protocol optimized to leverage fast networking~\cite{Derecho}. 
Each worker periodically multicasts updates about its local worker's state information to all other workers. However, we limit the frequency of these updates to ensure that overhead will be low. The staleness of the information a worker sees about other workers has an upper bound that is equal to the dissemination interval. In the experiments at Section~\ref{sec:Experiments and Evaluation}, we showed the scheduling mechanism has a high tolerance of information staleness, and enabled us to configure the update frequency to ensure that state information remains sufficiently accurate for Compass's task assignment decisions.


%% file: 4_compass.tex
\section{Compass Scheduler Design} 
\label{sec:CompassSchedulerDesign}

Compass employs a heuristic scheduling mechanism with dynamic coordination between workers. 
The mechanism exhibits an important difference from schedulers in prior literature in that Compass manages GPU memory and balances task assignment using a metric that reflects ML model object placements. The experiments reported in Section~\ref{sec:Experiments and Evaluation} confirm that this balance significantly improves end-to-end latency for job instances.

As explained in Section~\ref{sec:Task Dispatcher}, scheduling consists of two phases. The job instance planning phase produces the initial \dynamicgraph, and is described in Section~\ref{sec:planningPhase}.  The dynamic adjustment phase is presented in Section~\ref{sec:dynamic adjustment phase}.


\subsection{Parameters} 
\label{sec:parameters}
The Compass algorithm uses static information from the job \staticgraph, the \dynamicgraph data produced by the initial scheduler, information from the repository of workflow profiles, and real-time information about worker states provided by \statesMonitor.

\begin{description} [leftmargin=*]
\item[Task parameters] 
The expected runtime of task $t$ on worker $w$, $R(t,w)$, as well as the expected task input size $|\emph{input}_t|$ and output size $|\emph{output}_t|$ are obtained from the repository of workflow profiles.
The transfer duration $\emph{TD}_\emph{input}(t)$ for a task input is estimated via object size and network transmission capacity. The duration of transfer between two different workers is estimated by a commonly accepted heuristic as $\emph{TD}_\emph{input}(t) = |\emph{input}_t| / \emph{network transmision capacity} + \delta_\emph{network}$, where $\delta_\emph{network}$ is a constant factor contributing to the latency~\cite{data networks}.
This estimate is used when determining whether it would be more advantageous to schedule a successor to task $t$ on the same node, or to schedule it on a different node where the needed ML models will be local.  Tasks may also be executed on different nodes if scheduling a successor on the same node would force the latter task to wait because of resource oversubscription.

\item[ML model parameters] 
For ML model $m$, the model size $|m|$ is also stored in the repository of profiles. This information is used to estimate the time to fetch the model from host machine to GPU memory as follows: $\emph{TD}_\emph{model}(m,w) = |m| / \emph{PCIe transmission capacity}_w + \delta_{\emph{PCIe}}(w)$. It is also used to estimate the available memory that will remain in \CompassCache on the GPU after fetching the model.

\item[Worker parameters] 
The \statesMonitor collects and disseminates information about worker load and backlogs. Each worker $w$ estimates the time ($\emph{FT}(w)$) that would be required to fully execute all the tasks on its current \execqueue. This is computed as $\emph{FT}(w) = \emph{current time} + \sum_{t \in \execqueue} R(t, w)$.
We similarly define $\emph{FT}(t,w)$ as the estimated finish time of task $t \in \execqueue$ on worker $w$.
The available memory in the \CompassCache of worker $w$ is denoted $\emph{AVC}(w)$. 
It is derived from the cache size and the cumulative size of the models currently in the cache: $\emph{AVC}(w) = \emph{gpu capacity}_w - \sum_{m \in \text{the cache of } w} |m|$. 

\end{description}

\subsection{Planning Phase} 
\label{sec:planningPhase}
The task assignment algorithm used during Compass's job planning phase maps tasks in a job instance to workers. This algorithm is inspired by a well-researched process scheduling algorithm, Heterogeneous Earliest Finish Time (HEFT)~\cite{HEFT}. Like HEFT, Compass uses an upward ranking to prioritize the task scheduling order and selects the workers with the earliest start time. However, Compass extends the classic HEFT in several respects.  HEFT does not consider the load of a machine when performing the scheduling, which in our setting could result in inefficient task assignment.  Additionally, HEFT lacks mechanisms to factor in the potential benefits of co-location of the task with cached model objects on which it depends, or with input data. A third issue arises because HEFT locks down a plan for the entire job at the outset, preventing adjustments in the event that worker state changes in a significant way as the job instance proceeds. We address the first two limitations in the Compass's job planning algorithm and the third in its dynamic adjustment phase.

\subsubsection{Vertex Ranking}
Like HEFT, Compass starts by assigning a rank to each task $t$. We use $t \prec t'$ to denote that task $t$ is a direct predecessor of $t'$ in the \staticgraph. Since the target worker $w$ is unknown at the time of ranking, $R(t)$ is the average of $R(t,w)$ over the worker set.
The rank is defined recursively as
\begin{equation}
\label{eq:rank}
    rank(t) = R(t) + max_{t \prec t'} (\emph{TD}_{\text{output}}(t) + rank(t'))
\end{equation}

%
Using this ranking, tasks in the same job instance will be prioritized based on their dependencies and impact on the end-to-end latency. A task has a higher priority than a successor that depends on its input. Priorities of tasks with the same level of dependencies will be determined by their execution time and the size of any intermediate data objects that the task requires but are not locally available, because data transmission cost affects the end-to-end latency.
In our setting, we see heavy reuse of {\staticgraph}s, resulting in job instances containing identically ranked tasks.  In these cases, time of arrival determines the ranking.  

Notice that many aspects of task ranking can be performed statically.  Compass carries out these computations just once, when the {\staticgraph} is initially loaded, then saves the results into its repository.  Later, when dynamic inputs become available, the statically-computed rankings will merely need to be updated, not recomputed from scratch.

\subsubsection{Task Assignments}

The role of Compass's job instance planning phase is to produce an \dynamicgraph, which is a map from task ID to worker ID. The algorithm starts by considering the vertex rankings associated with the \staticgraph.  Compass's Scheduler module iterates through the tasks in descending order of priority, carrying out per-task scheduling decisions.

For a task $t$, this is done by looping through all worker nodes and selecting the worker with the earliest start time. For a given worker $w$, there are three main constituents, the transfer duration $\emph{TD}_\emph{model}(t, w)$, the finish time for all queued tasks $\emph{FT}(w)$, and the finish time for transferring all inputs of $t$, $\emph{FT}_\emph{allInputs}(t, w)$. 

$\emph{TD}_{model}(t, w)$ is the time for worker $w$ to load ML model $m$ for use by task $t$. Leveraging the data managed by the \statesMonitor, the scheduler can determine which GPUs currently hold decompressed instances of each of the models in the \CompassCache for all workers in the system.
If a task is assigned to a worker that will need to fetch an ML model, the required time is computed as
\begin{align}
&\emph{TD}_{\emph{model}}(t, w) =  \notag \\
& \begin{cases}
            0,\textbf{ if} \text{ $m_t \in$ \CompassCache on $w$} \\
            \emph{TD}_{\emph{model}}(m, w),\textbf{ if} \text{ $m_t \not\in$ cache} \land |m_t| \leq \emph{AVC}(w) \\
            \emph{TD}_{\emph{model}}(m, w)  + \emph{eviction penalty},\textbf{ otherwise}
\end{cases}
\label{eq:modelFetchTime}
\end{align}
%
%

\emph{Eviction penalty}. 
Consider some task $t$ that needs to be scheduled.  Compass could schedule $t$ on a worker $w$ that already holds needed model $m$, or could also expand the pool of workers by assigning to some other worker $w'$ where $m$ is not currently in cache. 
Naively, we should prioritize $w$ over $w'$ if the expected delay at $w$ is longer than the transfer delay for loading $m$ at $w'$ \-- but this overlooks the likelihood that some other model $m'$ will be evicted from cache by $w'$ to make room for $m$.  The penalty is intended to capture this effect.


$\emph{FT}(w)$ is the estimated time for worker $w$ to finish all tasks on its {\execqueue}. 
At each run of job instance planning, the scheduler creates a map of finish times for all workers, $\emph{worker}\_\emph{FT}\_\emph{map}$ by fetching the information from \statesMonitor,
as shown on line~\ref{declare_ft} of Algorithm~\ref{alg:PlanAlgo}. Scheduler also updates the map as it assigns tasks to workers (see Algorithm~\ref{alg:PlanAlgo}, line~\ref{update_ft}), since the assignments would affect later tasks in this job instance if scheduled onto the same worker. The planning algorithm incorporates $\emph{FT}(w)$ to account for the wait time on the workers' queue.

$\emph{AT}_\emph{allInput}(t, w)$ is the arrival time for all inputs of task $t$ to be received by worker $w$. It is calculated as the maximum among the arrival times on worker $w$, for the outputs produced by the predecessors of $t$. Because the scheduler processes the tasks according to their rankings, when task $t$ is examined its prerequisite tasks will have already been assigned. During job instance planning, Scheduler keeps track of its task assignments, and the estimated finish time of task $t$ on a scheduled worker $w$. 
Thus, the time when the output of task $t'$ arrives from worker $\dynamicgraph[t']$   at worker $w$ is computed as
\begin{equation}
    \emph{AT}_\emph{input}(t’, t, w) =
\begin{cases}
    \emph{FT}(t', \dynamicgraph[t']) ,\textbf{if } w = \dynamicgraph[t'] \\
    \emph{FT}(t', \dynamicgraph[t'])  + \emph{TD}_\emph{output}(t'), \text{otherwise}
\end{cases}
\end{equation}
\begin{equation}
    \emph{AT}_\emph{allInputs}(t, w)  = max_{t' \prec t}
    \emph{AT}_\emph{input}(t’, t, w)
\end{equation}
Algorithm~\ref{alg:PlanAlgo} shows the computation steps of the planning algorithm.
The complexity of the algorithm is $O(E*W)$, where $E$ is the total number of edges in the \staticgraph and $W$ is the total number of worker nodes.

\begin{algorithm}
\small 
\caption{Job Planning Algorithm}\label{alg:PlanAlgo}
\begin{algorithmic}[1]
\Require \staticgraph
\Ensure \dynamicgraph
\State compute the ranks for all tasks in the \staticgraph using Equation~\ref{eq:rank}
\State populate $\emph{worker\_FT\_map}$ from \statesMonitor
\Comment{ $\emph{worker\_FT\_map}$ is a map containing $\emph{FT}(w)$ $\forall w \in \emph{workers}$}
\label{declare_ft}
\State $\emph{TaskSet} \gets \emph{all tasks} \in \staticgraph$
\While {\emph{TaskSet} $\not= \emptyset$}
    \State $t \gets$ \text{a task in \emph{TaskSet} with the biggest rank}
    \State \emph{TaskSet} $\gets$ \emph{TaskSet} $\backslash t$
    \For {$w \in \emph{workers}$}
        \State
        $x \gets max( \emph{worker\_FT\_map[w]}, \emph{AT}_\emph{allInputs}(t, w) )$
        \State $ \emph{FT}(t, w) \gets x + \emph{TD}_\emph{model}(m_t) + R(t, w) $
        
    \EndFor
    \State $w_\emph{min} \gets \arg\min_{w \in \emph{workers}} \emph{FT}(t,w)$
    \State $\dynamicgraph[t] \gets w_\emph{min} $
    \State $\emph{worker\_FT\_map}[w_\emph{min}] \gets \emph{FT}(t, w_\emph{min})$ \label{update_ft}
\EndWhile
\State return \dynamicgraph
\end{algorithmic}
\end{algorithm}



\subsection{Dynamic Adjustment Phase} 
\label{sec:dynamic adjustment phase}

Whereas job planning occurs when a new job instance is created,  dynamic adjustment enables the system to adapt if predicted runtimes, object sizes, or transfer times were estimated inaccurately. 
Each time the Execution Engine finishes executing some task $t$, Scheduler will run  Algorithm~\ref{alg:AdjustAlgo}. It first checks that $t$ is not a join task: such a task cannot be moved to a different worker without coordination across the predecessor tasks.  For a non-join, it reschedules if the wait time on the planned worker exceeds a preconfigured threshold.

If rescheduling is needed, the new worker assignment is performed by ranking workers and selecting the one that would start the task first. 
The start time is calculated based on the wait time on the worker's \execqueue, required model fetching time, and the input transfer duration.

\begin{algorithm}
\small 
\caption{Task Dynamic Adjustment Algorithm}\label{alg:AdjustAlgo}
\begin{algorithmic}[1]
\Require $\emph{ADFG}, t$ 
\Ensure $\emph{opt\_worker}$
\State $w_\emph{planned} \gets \dynamicgraph[t] $
\label{check adfg}
\State $\emph{above\_threshold} \gets FT(w) > R(t,w) * \emph{threshold}$
\State $\emph{require\_reschedule} \gets (t \text{ is not a join task}) \land \emph{above\_threshold} $
\If{$\neg  \emph{require\_reschedule}$}
    \State return $w_\emph{planned}$
\EndIf
\State populate $\emph{worker\_FT\_map}$ from \statesMonitor
\For {$w \in \emph{workers}$}
    \State $x \gets \emph{worker\_FT\_map}[w]$
    \State $\emph{FT}(t, w)  \gets x + \emph{TD}_\emph{model}(t, w) + R(t,w) $
    \If{$w$ is not this scheduler's worker}
        \State $\emph{FT}(t, w)  \gets \emph{FT}(t, w) + \emph{TD}_\emph{input}(t) $
    \EndIf
    
\EndFor
\State $\emph{opt\_worker} \gets \arg\min_{w \in \emph{workers}} \emph{FT}(t,w)$
\State return $\emph{opt\_worker}$
\end{algorithmic}
\end{algorithm}

%% file: 5_implementation.tex
\section{Implementation}
Compass runs as a component of an open-source platform called Cascade~\cite{cascade,Derecho}, which was created to reduce overheads for ML jobs.  Prior to our work, Cascade held data (like a key-value store, file system, or message queuing middleware system).  It also offered a hosting environment for low-latency ML tasks.  Compass enriches this with ML scheduling.  

Cascade objects are simply variable-length byte vectors named by file-system pathnames. Each has a small set of home servers selected using a randomized hash-based object placement within "shards" of size 2 or 3.   Access is free on a home server, but a network transfer would occur for an access initiated from some other server.  ML models and other inputs to ML tasks would all be Cascade objects. 

Cascade additionally includes a storage/retrieval layer for in-memory replication of small data objects.  This layer is called the shared state table (SST), and is used by Compass to disseminate scheduler metadata (see Section~\ref{sec:sst}).

Lacking Compass, Cascade offered an automated hash-based load-balancer that was used in situations where a pool of servers all could handle a given request.  Integration of Compass with Cascade centered on replacing this mechanism with the decentralized Compass scheduler  and extending Cascade's preexisting support for GPU accelerators with Compass's GPU memory management functionality.  

\subsection{Data communication layer}
\label{sec:dataCommunicationLayer}
\subsubsection{Remote Direct Access Communication and the Data Plane Developers Kit}
\label{sec:rdma}
Cascade communication is optimized for modern networking, with a focus on two technologies receiving substantial industry attention: RDMA and DPDK.  Cascade supports a mix of communication modalities: point-to-point messaging, atomic multicast, and durable updates using a Paxos-based replication model.  For example, object replication within a shard uses atomic multicast or Paxos (the former for in-memory performance; the latter for persistence).  The SST data structure maps to point-to-point messaging.   Each of these, in turn, would be carried out using RDMA or DPDK, depending on which option Cascade is configured to select.

Remote Direct Memory Access (RDMA)~\cite{rdma} is a technology originally created for high-speed computing on bare-metal HPC clusters equipped with Infiniband networks.  During the past decade, RDMA has migrated to conventional data centers~\cite{RDMA1}.  The technology is fast, reliable, and offers TCP-like guarantees.  By moving the protocol stack into hardware, RDMA frees the host computer to focus on other tasks.  However, fully leveraging RDMA requires a substantial rethinking of end-to-end protocols and systems: its speed is closer to that of a computer backplane than a network. 

The Data Plane Developers Kit (DPDK) is a pure software solution, and runs over TCP.  DPDK moves the TCP stack to user space, bypassing the operating system kernel and by so doing, gaining a significant speedup.  In Cascade experiments, DPDK is about twice as fast as TCP and has much lower latencies, but RDMA offers a further factor of two in throughput and a further reduction in latency, particularly for data transfers larger than about 10KB.  
 
\subsubsection{Data Transfer between Task Dispatchers}
Although Compass's algorithms anticipate object transfers and include cost estimates, Compass itself does not currently initiate such actions.  Instead, the MLs themselves access objects by performing application-layer actions such as opening a file that is in fact hosted by Cascade, reading a key-value object hosted by Cascade, or receiving a pub-sub object that actually is a Cascade object.  In all of these cases, Cascade itself will then transparently fetch the object either from a local host cache, from local storage if the server making the access is a home node for the object, or from one of the home nodes if the object is hosted on other servers.  Compass does not currently try to "peek" into the Cascade host cache, but instead adopts the view that every object accessed during an ML job will be in memory "somewhere" in the system, and also that if the object in question is a large ML model \-- the case that can become costly \-- that DMA transfer from the local cache into GPU memory is comparable in speed to RDMA or DPDK transfer from a remote server's memory to the local GPU memory.  

\begin{figure}[h]
    \centering
    \includegraphics[width=0.8\columnwidth]{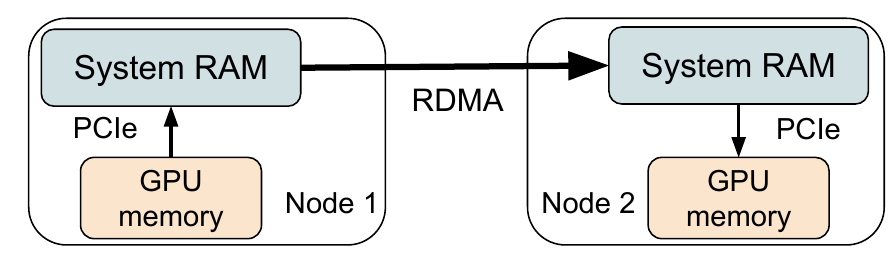}
    \caption{Network Transfer between Nodes}
    \label{fig:network transfer}
\end{figure}

Compass treats one kind of object differently: results of a task that become inputs to a successor task.  Here, the distinction is that whereas the outputs of some programs need to be saved and will be reused later, when we consider a {\staticgraph} there may also be transient outputs produced by one stage and then consumed and discarded by its successor stages.  Here, the cost depends on where the producer stage runs, and where the consumers run.  If an object is small, or if it is large but resides in the GPU cache on a node where the consumer will run, this cost is 0.  This form of co-location is clearly very beneficial.  Otherwise, Compass simply assumes that a DMA or RDMA transfer will occur, and models the cost accordingly.   The resulting model is presented in Figure~\ref{fig:network transfer}.

\subsection{Shared State Table Considerations}
\label{sec:sst}

\begin{figure}[t]
    \centering
    \includegraphics[width=0.9\columnwidth]{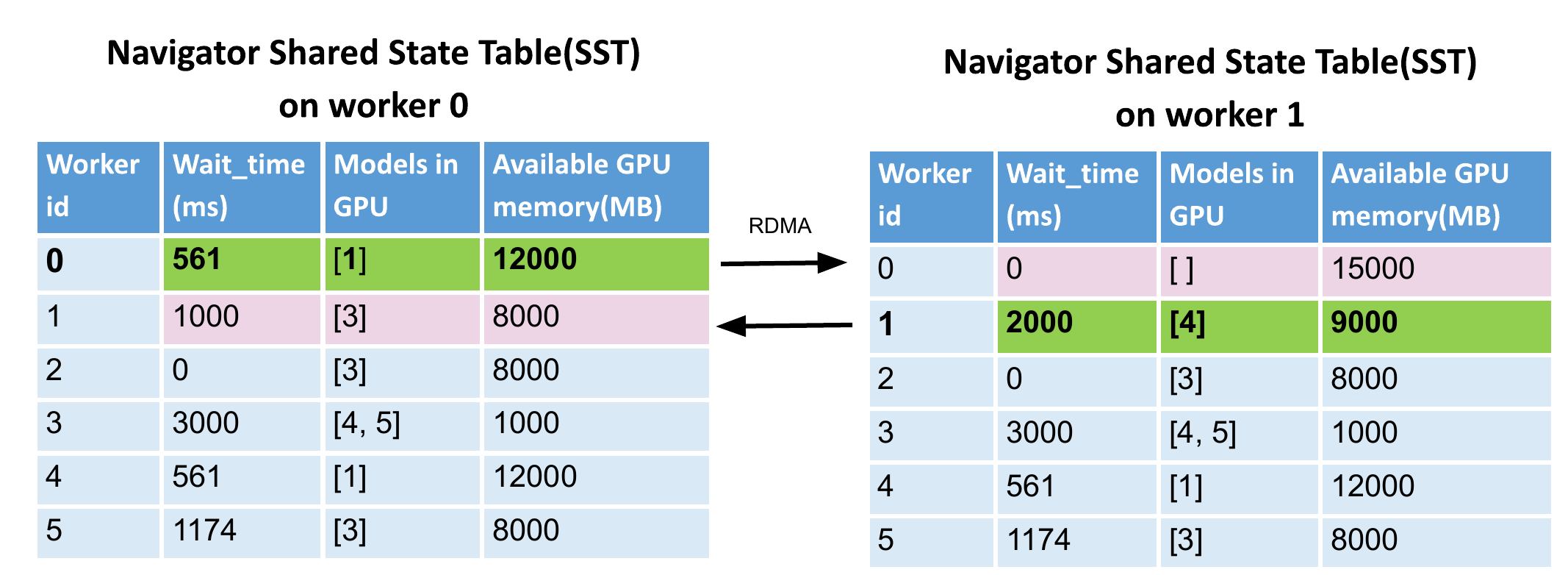}
    \caption{Compass Shared State Table (SST)}
    \label{fig:sst}
\end{figure}

While most objects stored in Cascade are simply key-value tuples with home nodes, but accessible from anywhere, the SST is a very different kind of storage structure.  As summarized above, the SST is an extensible table replicated over all servers.  Compass's "instance" of the SST has one row per server, and a fixed row format.  The resulting replicated object is an inherently $O(n^2)$ data structure: every node has a copy of the data structure, and when a node updates its row and {\em pushes} that update, one RDMA write per peer will be issued, to update the corresponding row in that peer.  Fortunately, $n$ in our target settings would generally be fairly small: most edge compute clusters have just a few (2 to perhaps 64) blade-style computers in a rack.  Even so, we limit the update rate to ensure that overhead is not excessive.

The nature of RDMA hardware makes SST updates {\em cache-line atomic}, meaning that data fitting with a 64-byte cache-line can be changed and the updates will become visible "instantaneously" when the push occurs.  However, data straddling cache lines would only be updated atomically on a per-cache-line basis.  This concern led us to squeeze the required data into a single cache-line per process, resulting in the SST layout seen in Figure~\ref{fig:sst}, for six nodes.  Our encoding uses a 64-bit integer to represent GPU cache contents, and hence is currently limited to 64 active models (we number them 0-63, and set that bit if the model is current resident in the GPU), but our design could easily be modified to allow 512 active models by expanding the encoding to use 2 cache lines rather than just one.

Compass's decentralized algorithm requires that a node making a scheduling choice be able to estimate the states of other nodes, but does not require perfect accuracy or lock-step coordination.  Accordingly, we decided to limit the frequency of SST updates using a parameter in the Compass configuration file, and then to experimentally determine the lowest acceptable update rate, thereby ensuring that SST overheads will be minimized.  In Section~\ref{sec:AlgorithmAnalysis} we report on these experiments, confirming that Compass performs well even with somewhat stale metadata, and justifying an update frequency of 5 pushes per second: negligible given that RDMA can send 75M small messages per second on each network link.

\subsection{GPU Memory Management}
\label{sec:gpuMemoryManagementPolicies}
Compass's GPU Memory Manager controls caching for recently used ML models in GPU memory. It performs model fetching and eviction as new tasks are assigned to the associated worker. Its fetching and eviction policy can be configured by users, depending on the application. We implemented and experimented with two management policies: FIFO and queue-lookahead.

\subsubsection{FIFO}
Under FIFO GPU memory management policy, GPU Memory Manager keeps a list of the models in GPU memory, sorted by the time when the model was added to the cache. 
When eviction is required to host a new model, cached models that are not actively in use get evicted starting from the earliest model on the list and until enough cache space is freed to hold the new model. \statesMonitor disseminates a bitmap for the cache to other workers in the system.

\subsubsection{Queue-Lookahead}
The FIFO policy has the drawback that the evicted model may soon be required by a subsequent task, resulting in a high number of evictions and fetch operations.
To account for this case, we implemented an alternative queue-lookahead mechanism. Since the \execqueue contains the tasks that are scheduled to be executed on the worker, this information can be used to implement a smarter cache management policy.

Upon eviction, GPU Memory Manager looks ahead into the {\execqueue}, examining some fixed number of future tasks and their required models, and giving higher priority to models that will be used sooner. Then it sorts the list of models in GPU memory based on the priority. If a model must be evicted to make room for the model of the current task, GPU Memory Manager evicts models from the lowest priority to the highest. This way Compass can avoid evicting models that are needed for the near-future tasks, and the associated overhead of fetching these models soon again.

\subsection{Simulation}
Although Compass is a fully working system, we wanted a way to explore scenarios beyond what we can reasonably deploy on our limited-scale testbed. Accordingly, we implemented an event-driven simulator [git repo].  The simulator itself is similar to that of Sparrow~\cite{Sparrow}. It models all elements of the environment described in Section~\ref{sec:System Architecture}, and supports the same request types as the real system, as illustrated in Figure~\ref{fig:pipelines}. When estimating network delays and execution times, the simulator uses values we measured in the real system for input/output sizes, the sizes of ML models for our workload, and the task duration. Compass's simulator models the task arrival, task waiting on the queue, task execution, and task dispatch as events, processed in the order of the simulated time. We validated the simulator against the real experiment using the same scale of 5 workers and the same workload, and observed that the performance difference between simulation and real experiment lies within 5\% of the median numeric values.

%% file: 6_experiments.tex
\section{Experiments and Evaluation} 
\label{sec:Experiments and Evaluation}


We evaluate Compass on a dedicated cluster, using 5 servers as the worker nodes and one as a client node. All workers are Dell PowerEdge R740 machines with dual Intel Xeon Gold 6242 processors and 192 GB memory. Each is additionally equipped with a Tesla T4 GPU having 16GB memory. The servers are connected by InfiniBand with RDMA-enabled 100Gbps network. Experiments in Section~\ref{sec:AlgorithmAnalysis} and~\ref{sec:sensitivityAnalysis} are evaluated on a real system. Experiment in Section~\ref{sec:scalabilityAnalysis} is performed using the simulator.

Our experiments focus on performance when scheduling multiple {\staticgraph}s on a shared set of workers, and use a mix of the four workflows from Figure~\ref{fig:pipelines}. On an idle system with ML models cached in GPU, the average completion times would range from 1 to 3 seconds, reflecting a dependency on relatively large models. In the experiment, the client node initiates a mix of concurrent requests. Text inputs to the translation~\ref{fig:translation} workflow and dialogue workflow~\ref{fig:dialogue} are randomly selected from the GLUE benchmark dataset~\cite{glue}. The image inputs for image reading pipeline~\ref{fig:readimg} and 3d perception pipeline~\ref{fig:3dperception} are sampled from the COCO dataset~\cite{coco}.

\subsection{Performance Evaluation Metrics} 
The metric slow down factor measures how close the execution of each job instance came to the (possibly unachievable) lower bound for end-to-end latency. End-to-end latency is computed from when the job instance  arrives at the system to the time when the last task in the job instance completes (in our examples, always shown as the "exit task"). Although the general problem of DAG-structured job scheduling is NP-complete, as mentioned in Section~\ref{sec:objectives}, we can easily compute the lower bound for a given job instance:
it is simply the time it takes to run the job instance with the maximum possible task parallelism and all inputs cached on GPU, and can be calculated from the {\staticgraph} by assuming zero data transfer delay. The slow down factor of a job instance $j$ is represented as
\[
\emph{slow\_down\_factor}_{j} = \frac{\emph{end\_to\_end\_latency}_{j}}{\emph{lower\_bound}_{j}} \geq 1
\]
In reality, schedulers may never be able to achieve the slow down factor of one. It is nevertheless a useful metric for comparing the performance of different schedulers.

\subsection{Scheduling Algorithm Analysis} 
\label{sec:AlgorithmAnalysis}
Two primary considerations arise when scheduling workflows with {\staticgraph} dependencies. One entails optimizing for dependencies within each job instance. For example, tasks that could be processed in parallel can be sent to different workers to maximize the level of concurrency during execution; tasks with large intermediate data dependencies can be grouped to run on a single worker to avoid data transfer delays. The other entails optimization reflecting the current state of the system as a whole, as represented in the shared metadata table.  Here, decisions will avoid overloading a worker that is already at its peak capacity.  Both approaches offer potential benefits, but they lead to very different schedules.  The Compass algorithm takes both factors into consideration, using a hybrid scheduling scheme. Below in Section~\ref{sec:competingSchemes} we describe two scheduling schemes that each focus on one factor.  Then Section~\ref{sec:comparisonExperiment} compares Compass with these two baseline options.

\subsubsection{Baseline Schemes}
\label{sec:competingSchemes}
\begin{description} [leftmargin=*]
\item[JIT:] A Just-in-time (JIT) scheduler makes individual task assignment decision as each task is about to be executed. Each time a scheduling action is needed, a JIT algorithm selects a task that has all prerequisite inputs available and assigns it to the worker that offers the earliest start time, obtaining the start time estimates by taking worker-state information from {\statesMonitor} and computing the worker start time using the worker wait time, model fetch time and intermediate data transfer time (if a required input would need to be moved from a different worker). The JIT optimization minimizes the finish time for each individual task.

\item[HEFT:] The HEFT algorithm introduced in Section~\ref{sec:planningPhase} optimizes for inter-job dependencies. Similarly to Algorithm~\ref{alg:PlanAlgo}, it first sorts tasks in the job instance, then assigns tasks to workers in the descending order of priorities. The optimization considers task parallelism and data transfer between dependent tasks. Unlike the  customized planning Algorithm~\ref{alg:PlanAlgo}, a standard HEFT algorithm does not consider the worker wait time and the task-dependent ML model locality. It also lacks a dynamic adjustment mechanism: When a job instance is first triggered, HEFT would assign all the associated tasks to workers, and the workers subsequently adhere to the schedule.

\item[Hash:] The algorithm balances the load by randomizing task assignment to workers by hashing the task name combined with the request identifier.
Hash is very simple and offers a uniform distribution of tasks to workers, and it is commonly used for workflow scheduling and load balancing.
\end{description}

\begin{figure*}
    \centering
     \begin{subfigure}[b]{0.393\textwidth}
     \includegraphics[width=0.9\textwidth]{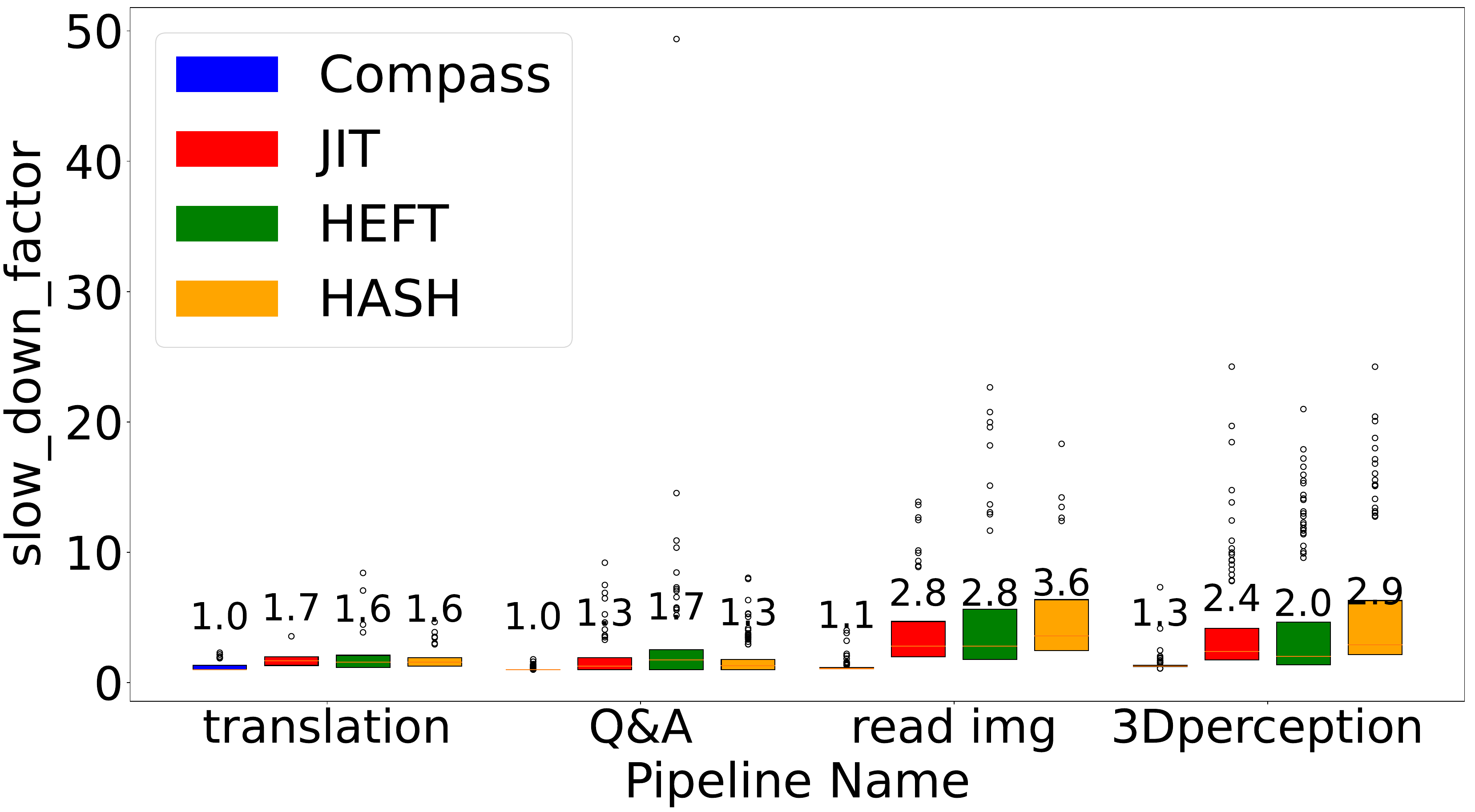}
     \caption{Low-load scheduling performance}
     \label{fig:comparison_low}
     \end{subfigure}
     \hspace*{-2.5em}
     \begin{subfigure}[b]{0.3836\textwidth}
     \includegraphics[width=0.9\textwidth]{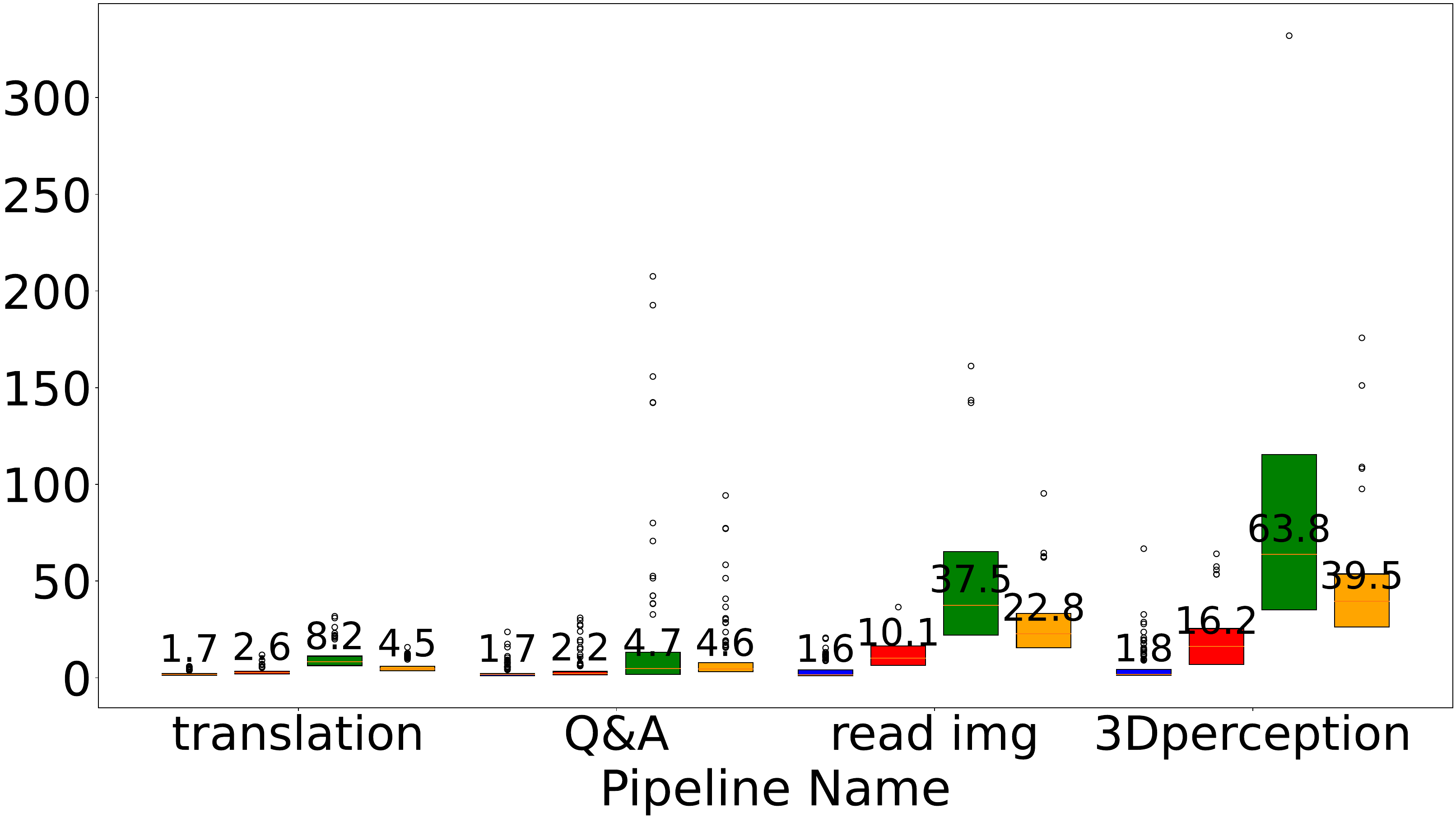}
     \caption{High-load scheduling performance}
     \label{fig:comparison_high}
     \end{subfigure}
     \hspace*{-2.3em}
     \begin{subfigure}[b]{0.276\textwidth}
     \includegraphics[width=0.9\textwidth]{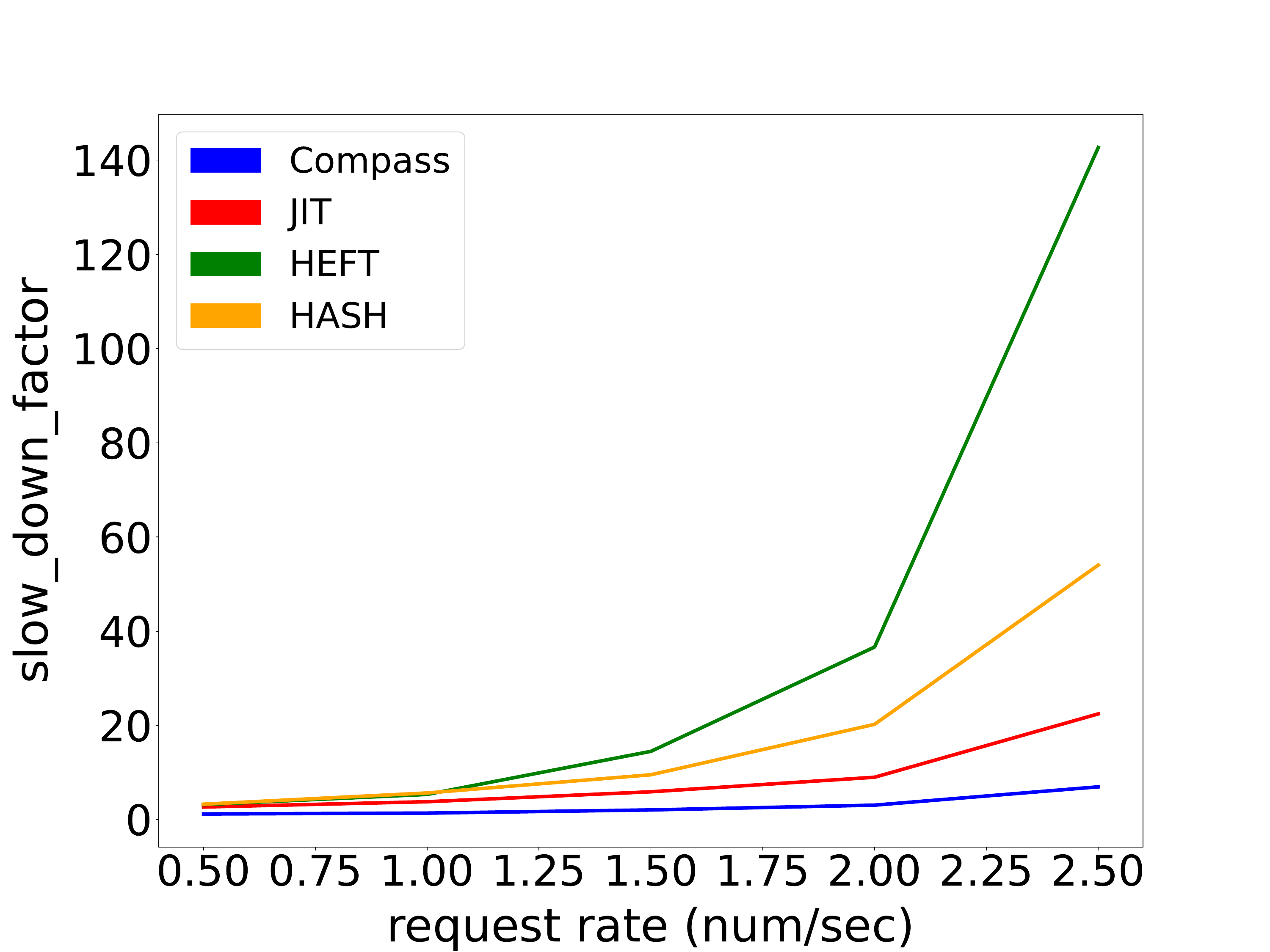}
     \caption{Varying load performance}
     \label{fig:comparison_vary}
     \end{subfigure}

\caption{Comparison of scheduling schemes}
\label{fig:schemes_comparison_experiment}
\end{figure*}

\subsubsection{Performance Comparison Among Schedulers}
\label{sec:comparisonExperiment}
Figure~\ref{fig:schemes_comparison_experiment} shows results for an experiment that compares the scheduling schemes under steady low and high workloads, using mixes of the four job types presented in Figure~\ref{fig:pipelines}.

Figure~\ref{fig:comparison_low} shows the low-load case. Clients send pipeline processing requests at the average rate of 0.5 requests per second, with a Poison distribution on request types.
The box plot shows the distribution of slow\_down\_factors for different instances, broken down by job category. The top and bottom of the box represent the first and third quartile of the data, respectively. The whiskers represent 1.5 times the upper and lower quartiles, and the dots show outliers. While all schedulers perform well under this case with close to optimal slow\_down\_factor, Compass is the closest to the optimal schedule (slow\_down\_factor of 1.0). 

Figure~\ref{fig:comparison_high} repeats the experiment with a job  arrival rate having a mean of 2 requests per second under a Poison distribution. This places the cluster under more pressure, so the slow\_down\_factor for all four scheduling schemes increases due to the higher load. Compass continues to outperform, yielding overall performance 2x to 4x faster than HEFT and Hash for the translation and Q\&A pipelines, and 20x to 30x faster for the image description  and 3D perception pipeline. The more extreme slow\_down\_factors in image description and 3D perception pipelines are because of their relatively short runtimes compared to translation and Q\&A, which made these pipelines more susceptible to the system overheads caused by sub-optimal schedules. 

Figure~\ref{fig:comparison_vary} experiments with different request rates under a Poison distribution. The plot shows the average slow\_down\_factors for a mixed workload. Compass has the closest to ideal performance at different request rates. The HEFT algorithm fails to adjust to the higher system loads due to not taking the workers' load into consideration. JIT performs better than the HEFT and Hash schemes due to its in-time task assignment mechanism, but not as well as Compass because of it lacks a mechanism for intra-job coordination.

Table~\ref{tab:scheduler_gpu_consumption_comparison} shows average latency and other GPU-related performance metrics for the experiment presented in Figure~\ref{fig:comparison_high}.
GPU utilization is the percentage of time during which the GPU was actively processing computations. Memory utilization is the percentage of GPU memory being used, an indicator of memory-pressure  on the GPU. Energy consumption is the total amount of energy consumed when running the full workloads. The GPU cache hit rate is the percentage of times when the required model is already in the GPU memory.
The results indicate that Compass consumes similar GPU resources and energy to other schedulers, yet achieves superior end-to-end latency.

\begin{table}[h]
\caption{Scheduler performance metrics}
\label{tab:scheduler_gpu_consumption_comparison}
\centering
{\footnotesize
\setlength\tabcolsep{2pt}
\begin{tabular}{|p{1.1cm}|p{0.8cm}|p{1.35cm}|p{1.6cm}|p{1.4cm}|p{1.3cm}|}
\hline
Scheduler & latency (s)& GPU utilization (\%) & GPU memory utilization (\%) & GPU energy  use (J) & GPU cache hit rate (\%)  \\
\hline
Compass & 2.5 & 39 & 65 & 100760 & 99 \\
\hline
JIT & 5.0 & 42 & 68 & 103254 & 93 \\
\hline
HEFT & 18.0 & 38 & 73 & 103475 & 95 \\
\hline
HASH & 10.5 & 39 & 67 & 106851 & 91\\
\hline
\end{tabular}
}
\end{table}

\subsection{Ablation Analysis}
Compass employs multiple features that contribute to scheduling. To understand the importance of each, we conduct an {\em ablation analysis} by selectively disabling features. 

\subsubsection{Importance of Scheduler Features}
\label{sec:AblationFeatureAnalysis}

\begin{figure}[h]
    \centering
    \includegraphics[width=0.9\columnwidth]{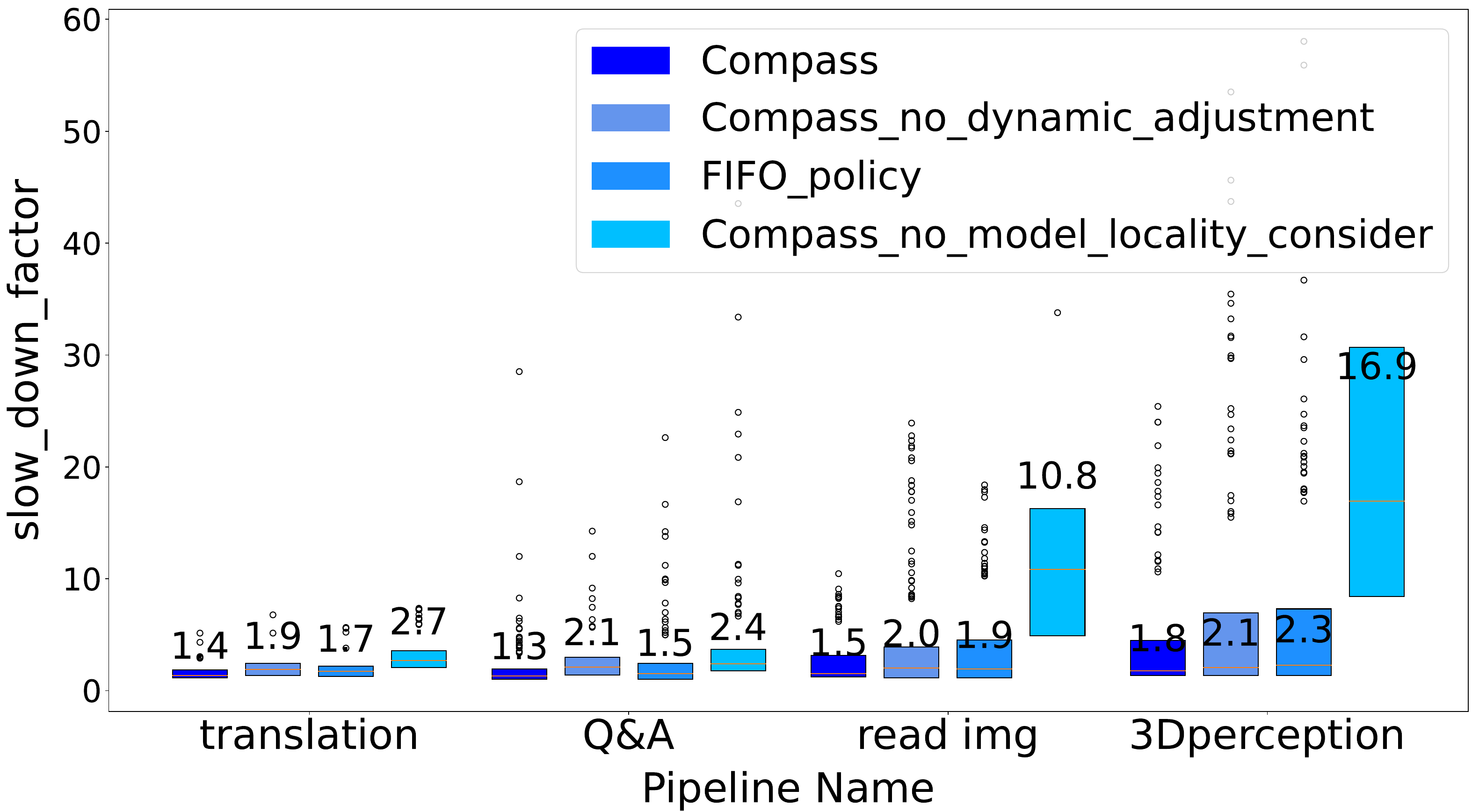}
    \caption{Ablation analysis findings}
    \label{fig:optimization_effect}
\end{figure}

\begin{description} [leftmargin=*]
\item[Dynamic task scheduling:]
To show the effect of using dynamic adjustment, this experiment disables
dynamic adjustment that Compass normally employs as each job instance executes, only retaining the preliminary planning phase. Figure~\ref{fig:optimization_effect} reveals a significant performance degradation. 
\item[Eviction policy:] 
Next we investigate the importance of GPU cache-content awareness by selectively disabling the Memory Management policies introduced in Section~\ref{sec:gpuMemoryManagementPolicies}. As seen in Figure~\ref{fig:optimization_effect},  queue lookahead improves latency performance when the request rate is high, but has no significant impact at low request rates, where the initial task  assignment decision is already close to optimal. 
\item[Model locality:]
The scheduler prioritizes workers whose GPU cache already contains the required models, as reflected in the metadata table. As shown in Figure~\ref{fig:optimization_effect}, Compass incurs an 8x degradation without this mechanism. The GPU cache hit rate also decreases to 90\% (as opposed to 99\% when this optimization is enabled).
\end{description}

\subsubsection{Sensitivity Analysis} 
\label{sec:sensitivityAnalysis}
Recall from Section~\ref{sec:System Architecture} that we made a decision to limit the update rate for the SST-hosted metadata on which Compass depends, forcing the system to run with potentially stale data.  The question arises of how sensitive Compass's scheduling quality actually is as a function of the degree of staleness.

\begin{figure}[t]
    \centering
    \vspace{40pt}
    \includegraphics[width=0.9\columnwidth]{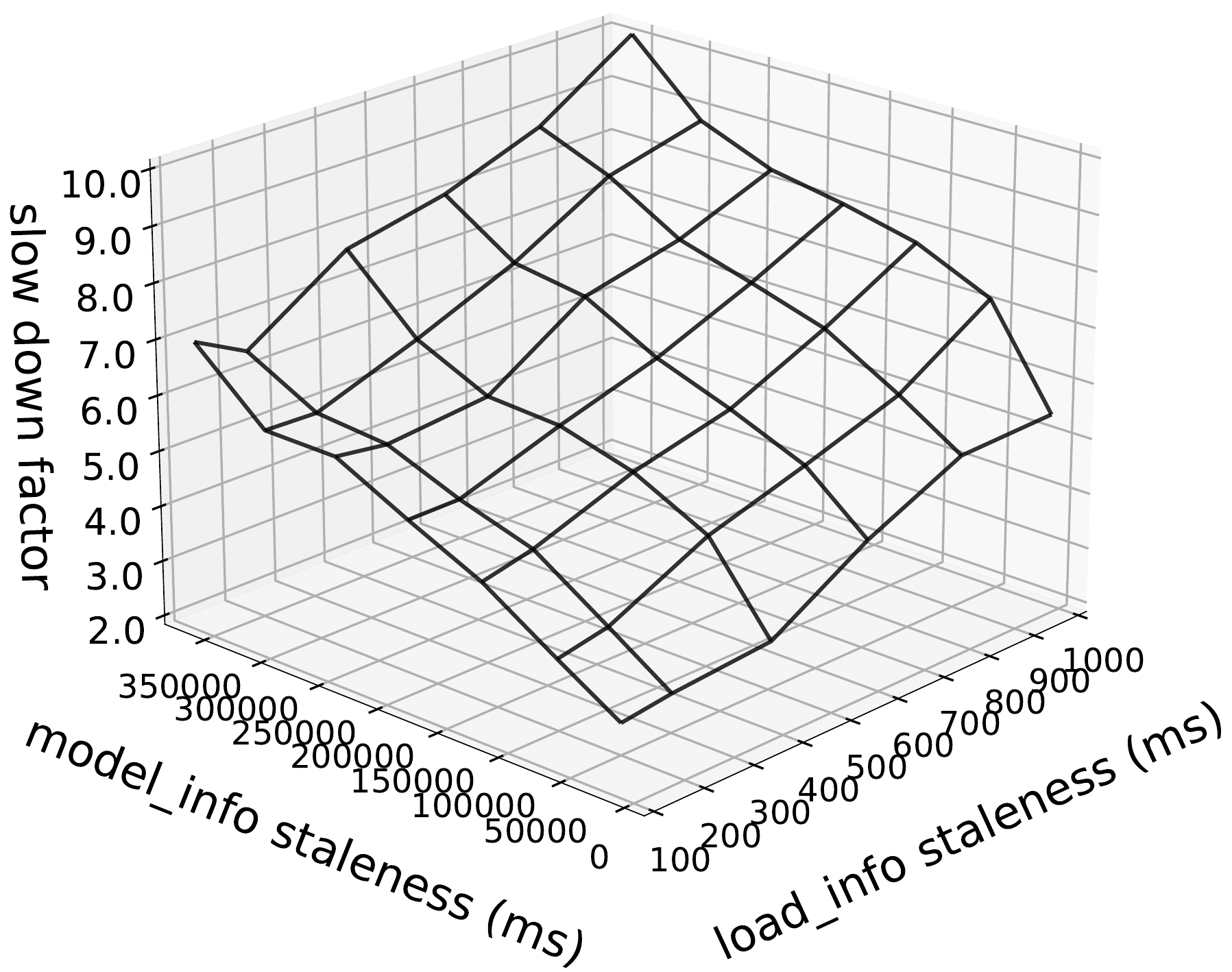}
    \caption{Dependency on information dissemination rate}
    \label{fig:vary_info_staleness}
\end{figure}

With this in mind, we set up a high-workload scenario and vary the frequency of updates using a "delay between updates" configuration parameter, reflected in Figure~\ref{fig:vary_info_staleness}. In this plot, the x axis represents the staleness of the workers' load information as perceived by other workers, whereas the y axis represents the staleness of the workers' GPU cached model information as perceived by other workers.  Observe first that the minimal slow\_down\_factor is achieved with the highest rate of SST pushes (10 per second), and the worst -- with a very low rate (1 per second).

The graph tells us quite a bit more, however.  Because ML model fetching to the GPU memory is an infrequent event, we see that Compass can tolerate a much higher level of staleness for GPU cache-content information than for load. The delay for the latter has a significant impact beyond the threshold of 200ms (corresponding to 5 pushes per second).  To make this concrete, at the 200ms rate in a cluster of 32 nodes, each participant would send and receive 160 RDMA updates per second (out of an RDMA rate limit of 75M/second): a trivially low overhead.

\subsection{Production trace}
\begin{figure}[ht]
     \centering
     \begin{subfigure}[b]{0.22\textwidth}
         \centering
         \includegraphics[width=\textwidth]{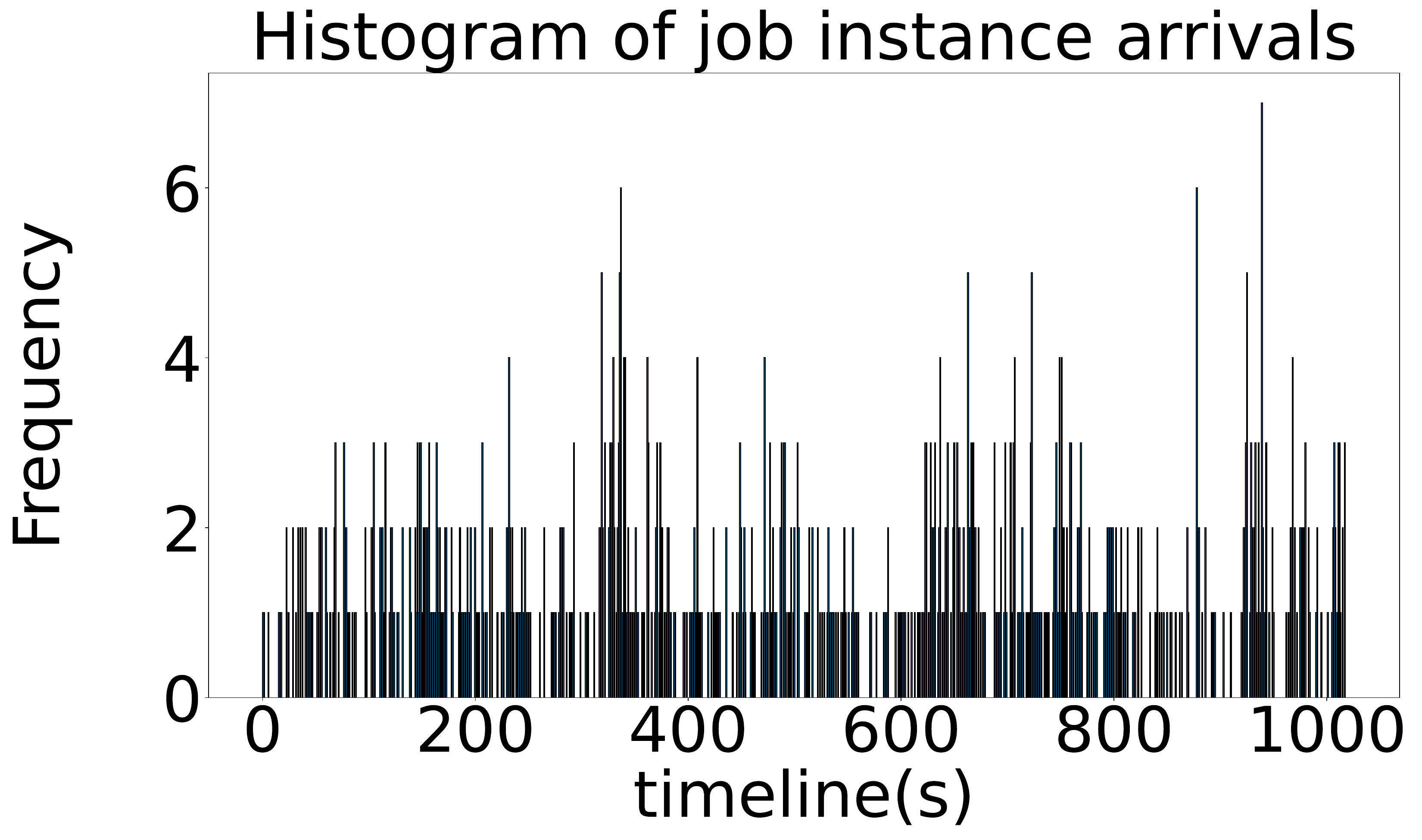}
         \caption{trace histogram}
         \label{fig:real_trace}
     \end{subfigure}
     \hfill
     \begin{subfigure}[b]{0.22\textwidth}
         \centering
         \includegraphics[width=\textwidth]{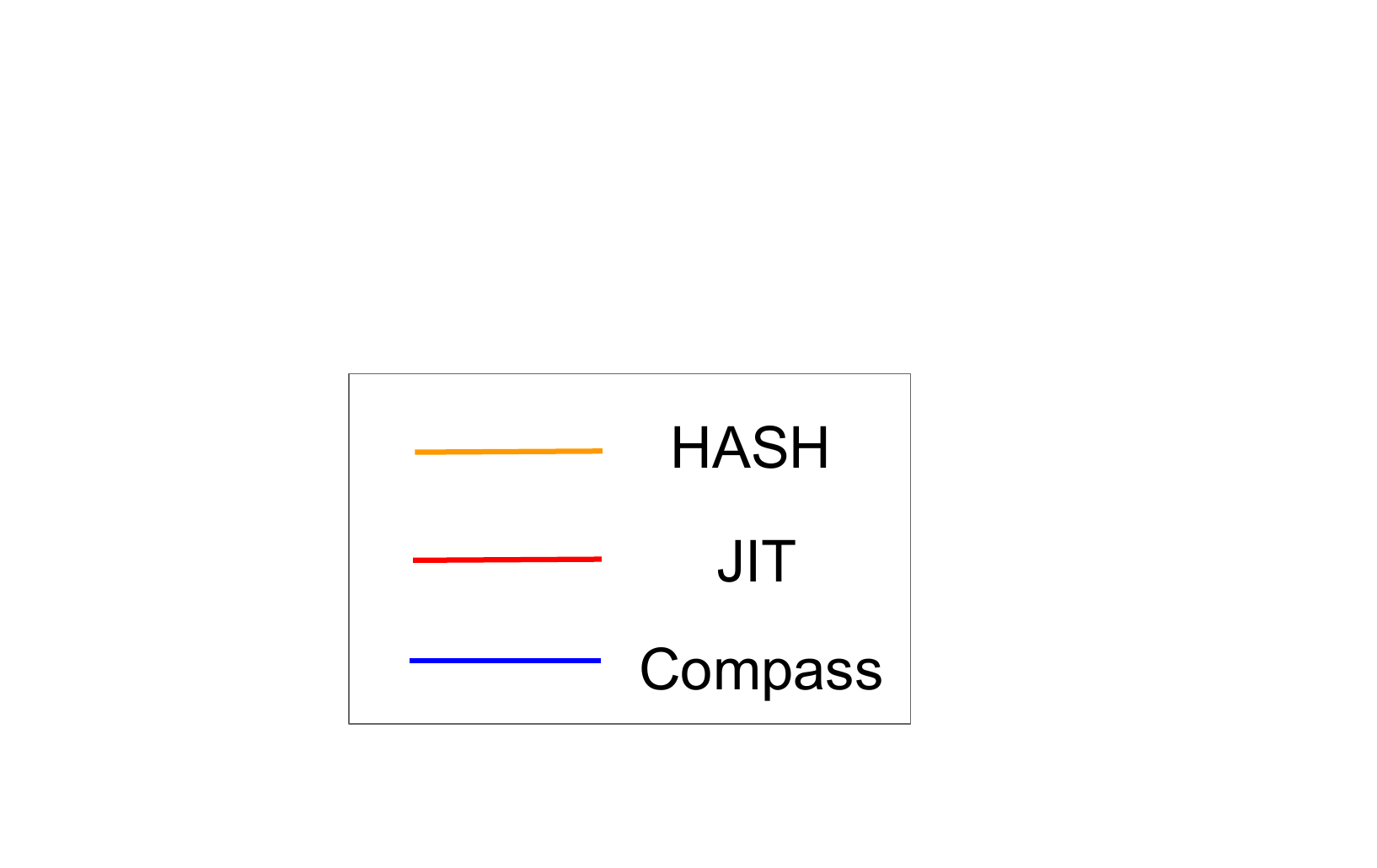}
         \label{fig:legend}
     \end{subfigure}
     \hfill
     \begin{subfigure}[b]{0.22\textwidth}
         \centering
        \includegraphics[width=\textwidth]{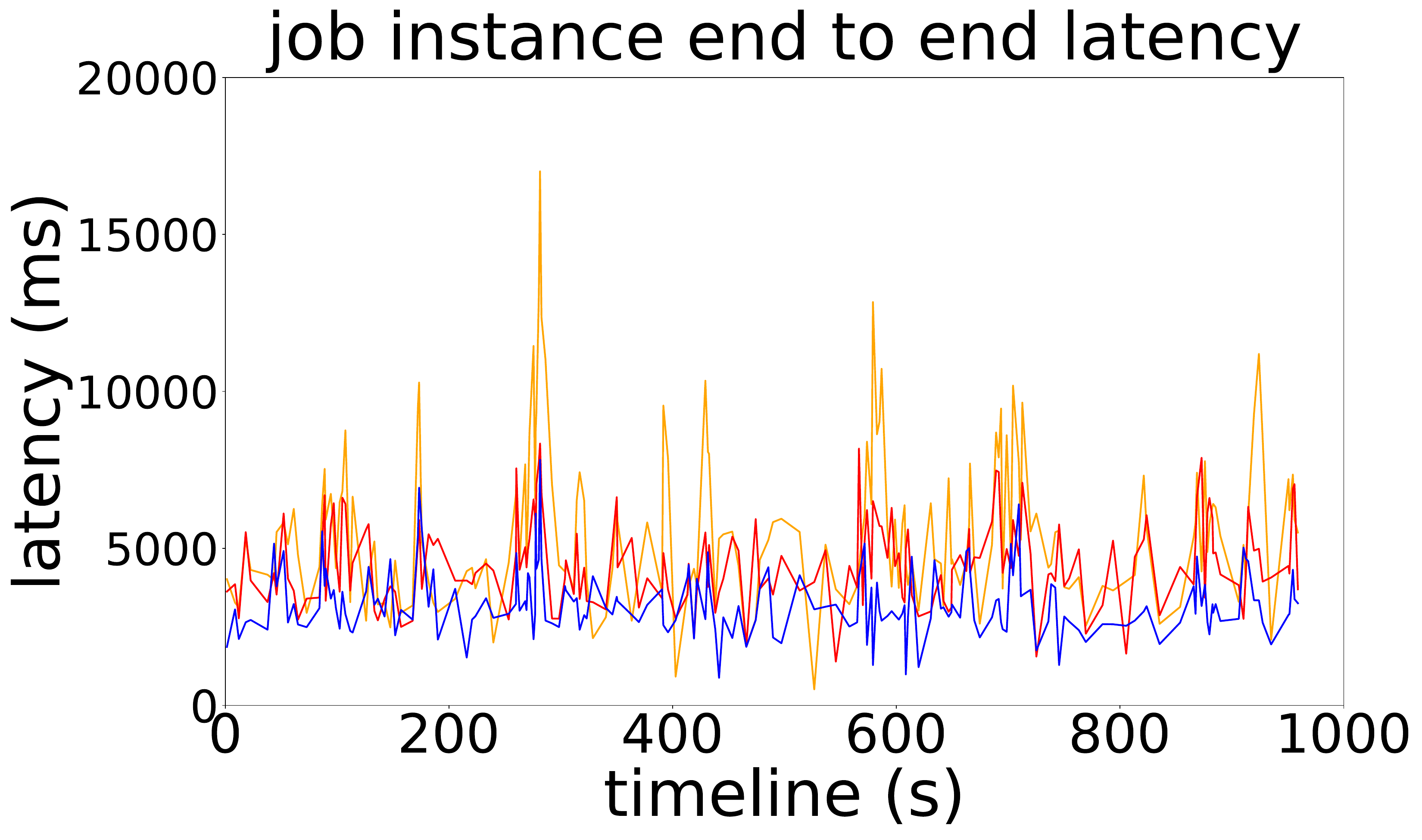}
         \caption{language translation}
        \label{fig:language_trace_experiment}
     \end{subfigure}
     \hspace*{-0.5em}
     \begin{subfigure}[b]{0.22\textwidth}
         \centering
         \includegraphics[width=\textwidth]{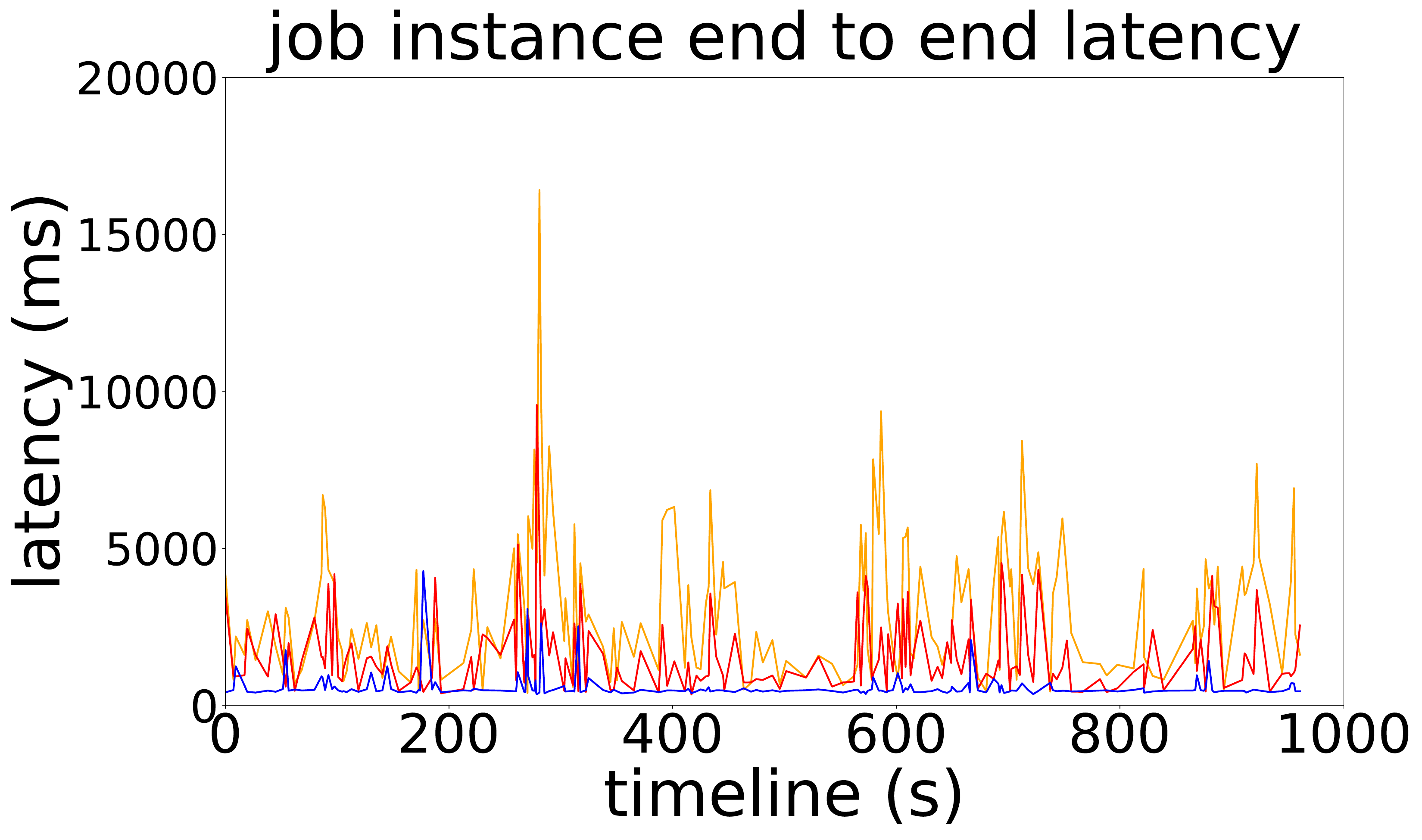}
         \caption{image reading}
         \label{fig:img_read_trace_experiment}
     \end{subfigure}
     \hfill
     \begin{subfigure}[b]{0.22\textwidth}
         \centering
         \includegraphics[width=\textwidth]{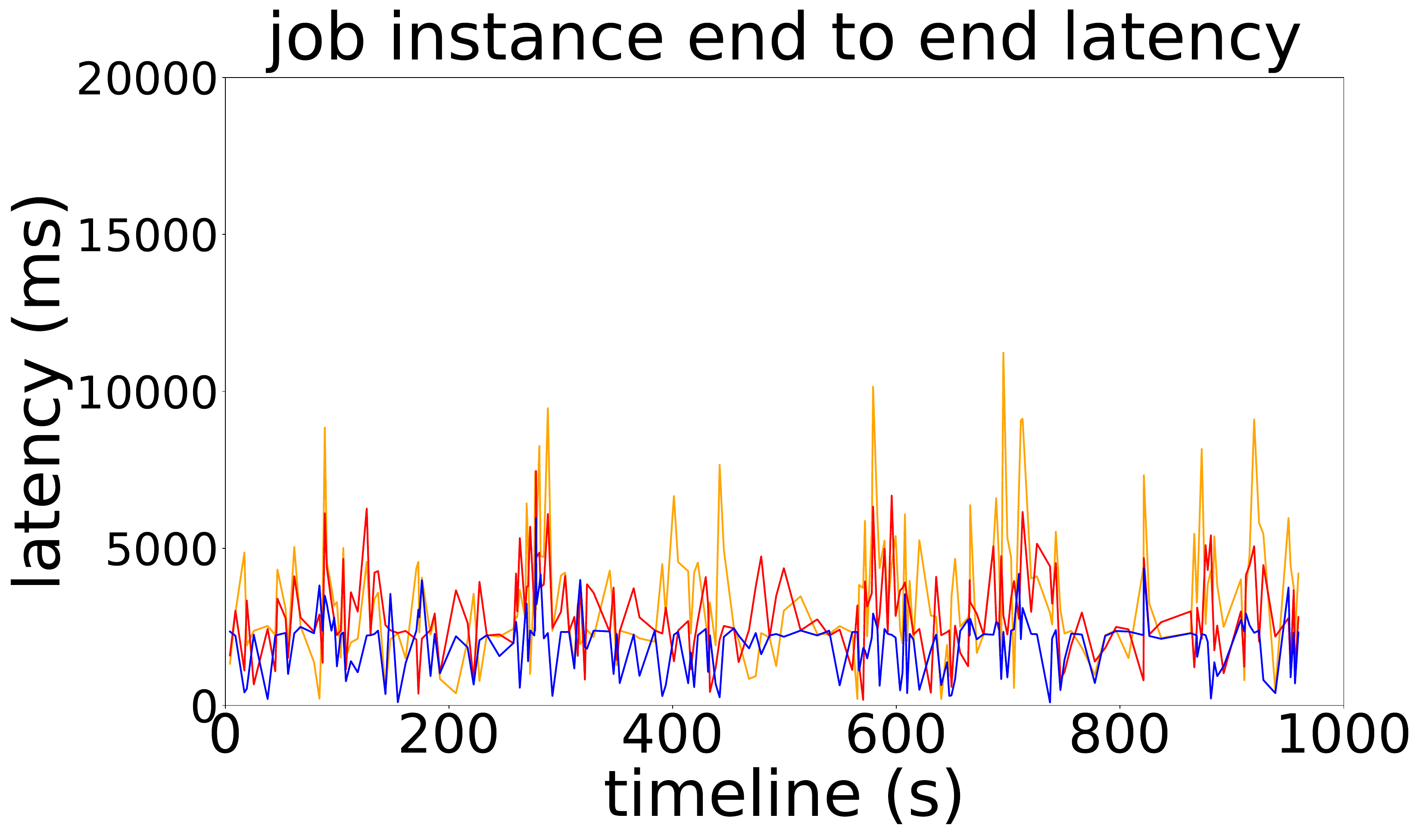}
         \caption{dialogue}
         \label{fig:question_answer_trace_experiment}
     \end{subfigure}
     \hspace*{-0.5em}
     \begin{subfigure}[b]{0.22\textwidth}
         \centering
         \includegraphics[width=\textwidth]{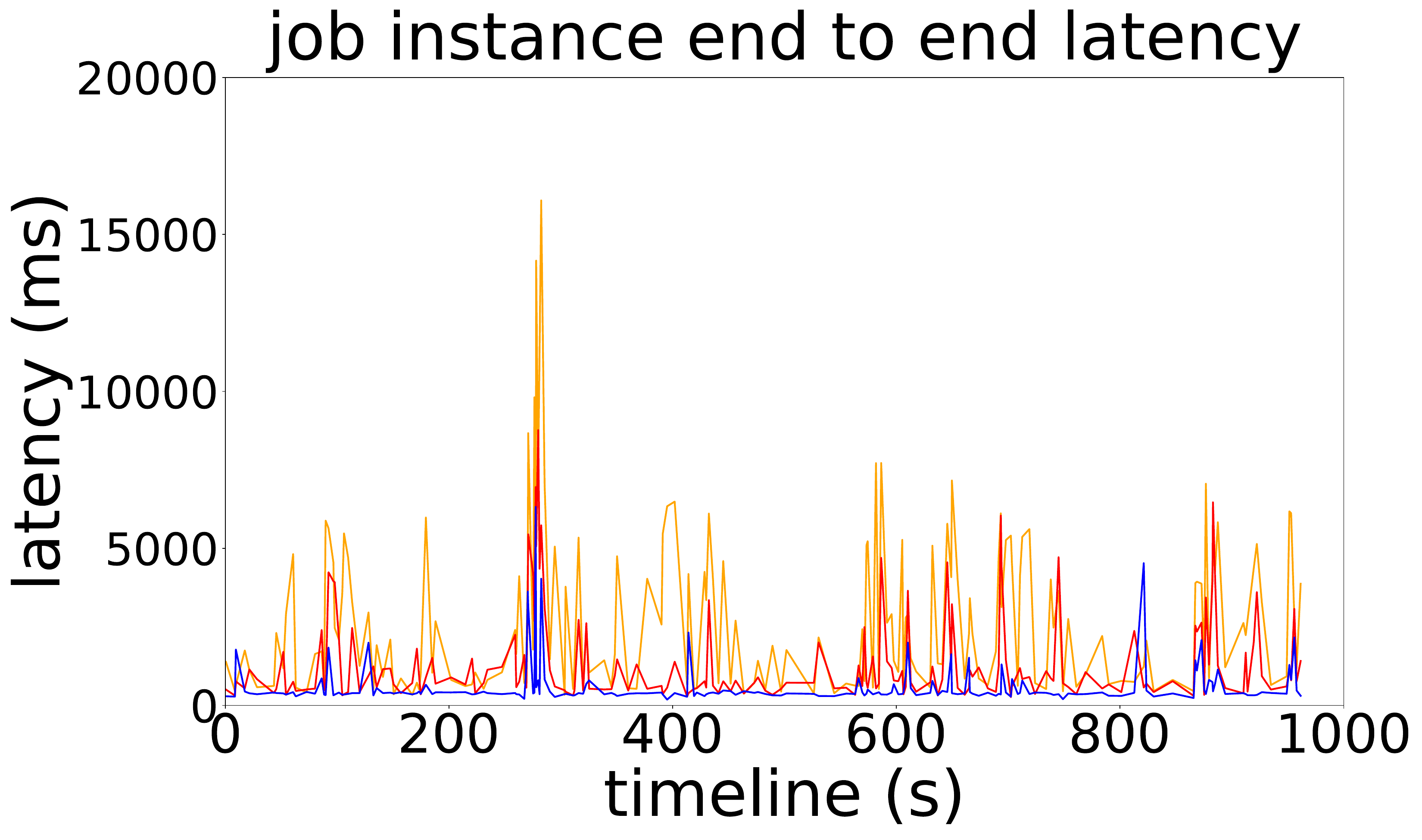}
         \caption{3d perception}
         \label{fig:3d_perception_trace_experiment}
     \end{subfigure}
\caption{Experiment on production traces}
\label{fig:real_trace_experiment}
\end{figure}

Our next experiment considers scheduling performance for the workflows described in Figure~\ref{fig:pipelines} using a public trace from the Alibaba production GPU cluster~\cite{cluster_data,mlaas_in_the_wild, characteristicsJobMicroarchitecture}. We rescale this trace to adapt it to the capacity of our experimental system. 
Figure~\ref{fig:real_trace} shows request arrival rates on a timeline, while Figures~\ref{fig:language_trace_experiment}-\ref{fig:3d_perception_trace_experiment} show completion times as a function of arrival times. The plots make it clear that the Hash scheduling scheme is least tolerant of bursts of job instances, at least in a small worker cluster. Compass has the best completion times even with these unpredictable request patterns.

\subsection{Scalability Analysis} 
\label{sec:scalabilityAnalysis}
Hashing is widely used for scalability in today's large clusters~\cite{energyEfficientWorkflowScheduling, multiObjectiveOptimization, hashKV}, under the assumption that it leads to a more even distribution of workloads and full use of resources.  Our next experiment shows that in fact, Compass uses hardware resources more efficiently. For this work we create a large-scale experiment using a simulation study.  The workload is Poisson with a mean of 40 requests per second. We consider a range of up to 250 workers and examine a series of cluster utilization metrics as well as the median slow\_down\_factor over the job instances of all four types in Figure~\ref{fig:pipelines}.

\begin{figure}[b]
    \centering
    \includegraphics[width=0.8\columnwidth]{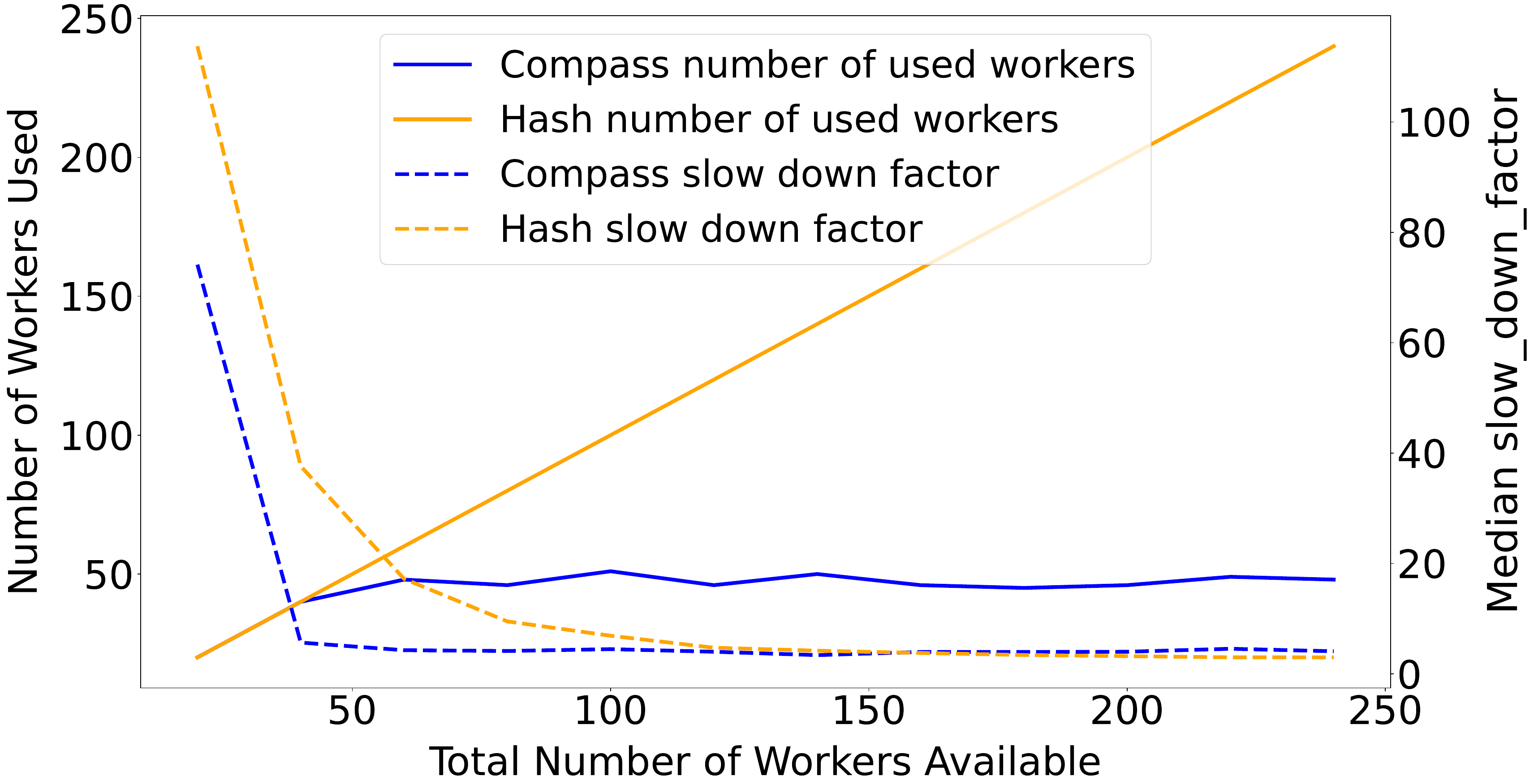}
    \caption{Simulation testing a larger scale}
    \label{fig:scalability_experiment}
\end{figure}

As seen in Figure~\ref{fig:scalability_experiment}, Hashing is effective in the sense that it utilizes all available workers in each round, and the median slow\_down\_factor decreases as the number of workers increases, due to its inherent load-balancing property, approaching its lower bound at the scale of around 100 workers. Compass prioritizes the workers with models in GPU memory and expands the worker set only if by doing so, mean job completion times would be improved.  As observed in the experiment with the same workload, the slow\_down\_factor reaches its lower bound with just 50 active workers: half the resources required for the Hashing scheduler! These unused machines could be put into power-saving mode, reducing overall platform power consumption.  With very large numbers of workers, Hashing slightly outperforms Compass (at 150 machines and beyond), but to gain this tiny benefit it keeps three times as many machines active.

\remove{

\begin{figure}[ht]
     \centering
     \begin{subfigure}[b]{0.23\textwidth}
         \centering
         \includegraphics[width=\textwidth]{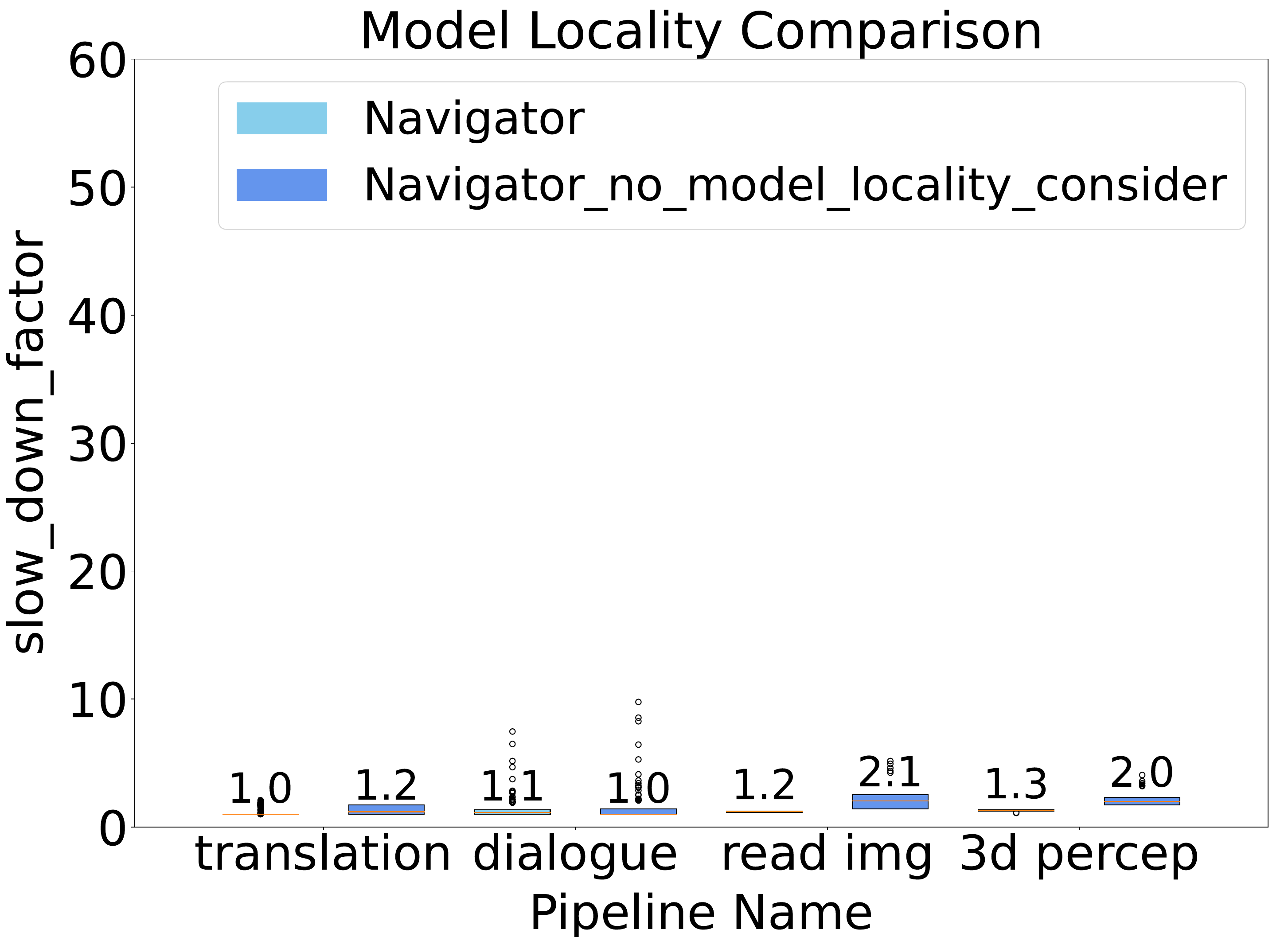}
         \caption{Model locality considered vs. not considered schemes comparison under low request rate}
         \label{fig:model_loc_low_rate}
     \end{subfigure}
     \hfill
     \begin{subfigure}[b]{0.23\textwidth}
         \centering
         \includegraphics[width=\textwidth]{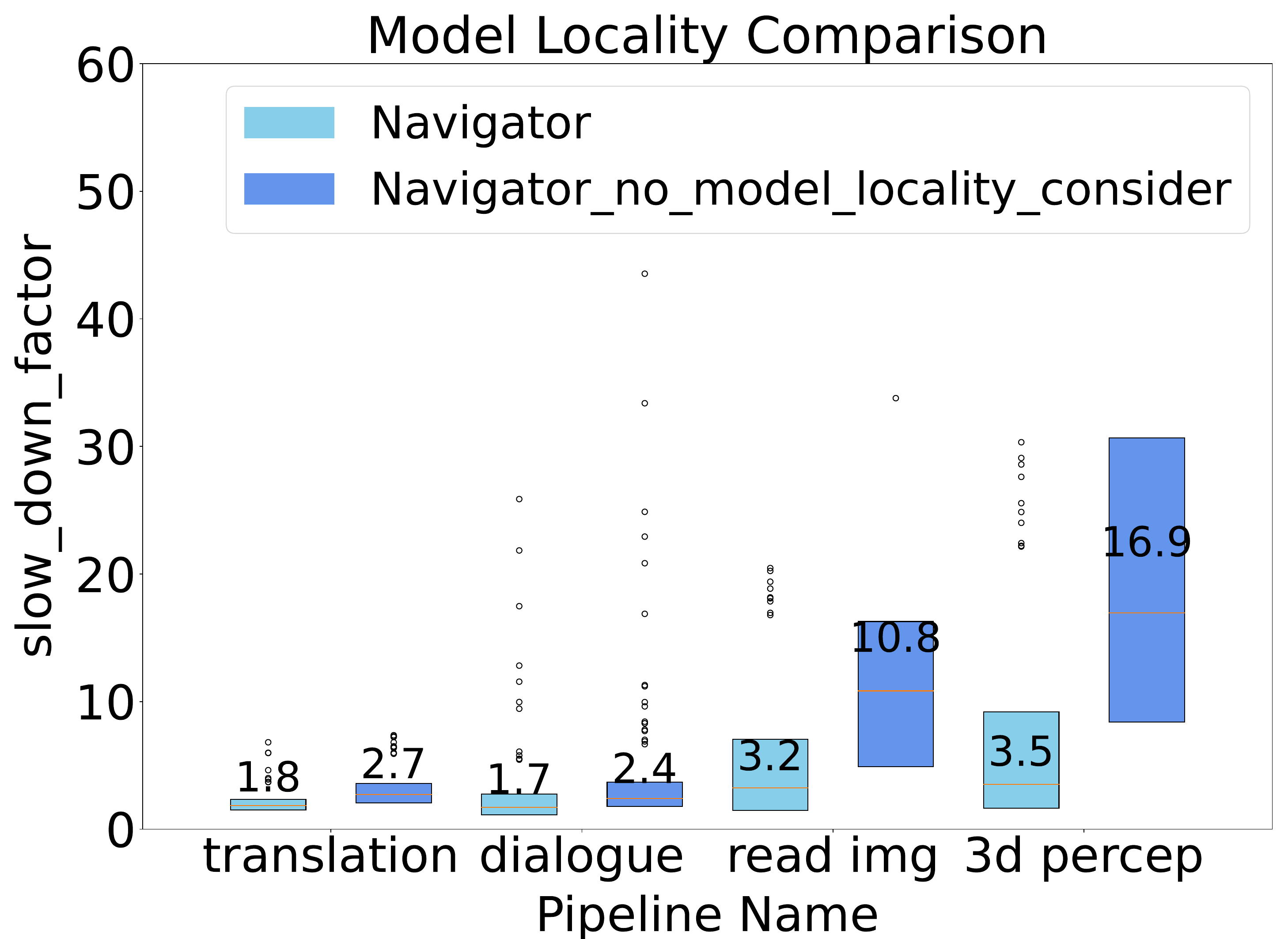}
         \caption{Model locality considered vs not considered schemes comparison under high request rate}
         \label{fig:model_loc_high_rate}
     \end{subfigure}
\caption{Model locality considered vs not considered schemes comparison}
\label{fig:memory_policy_comparison}
\end{figure}

\begin{figure}[ht]
     \centering
     \begin{subfigure}[b]{0.23\textwidth}
         \centering
         \includegraphics[width=\textwidth]{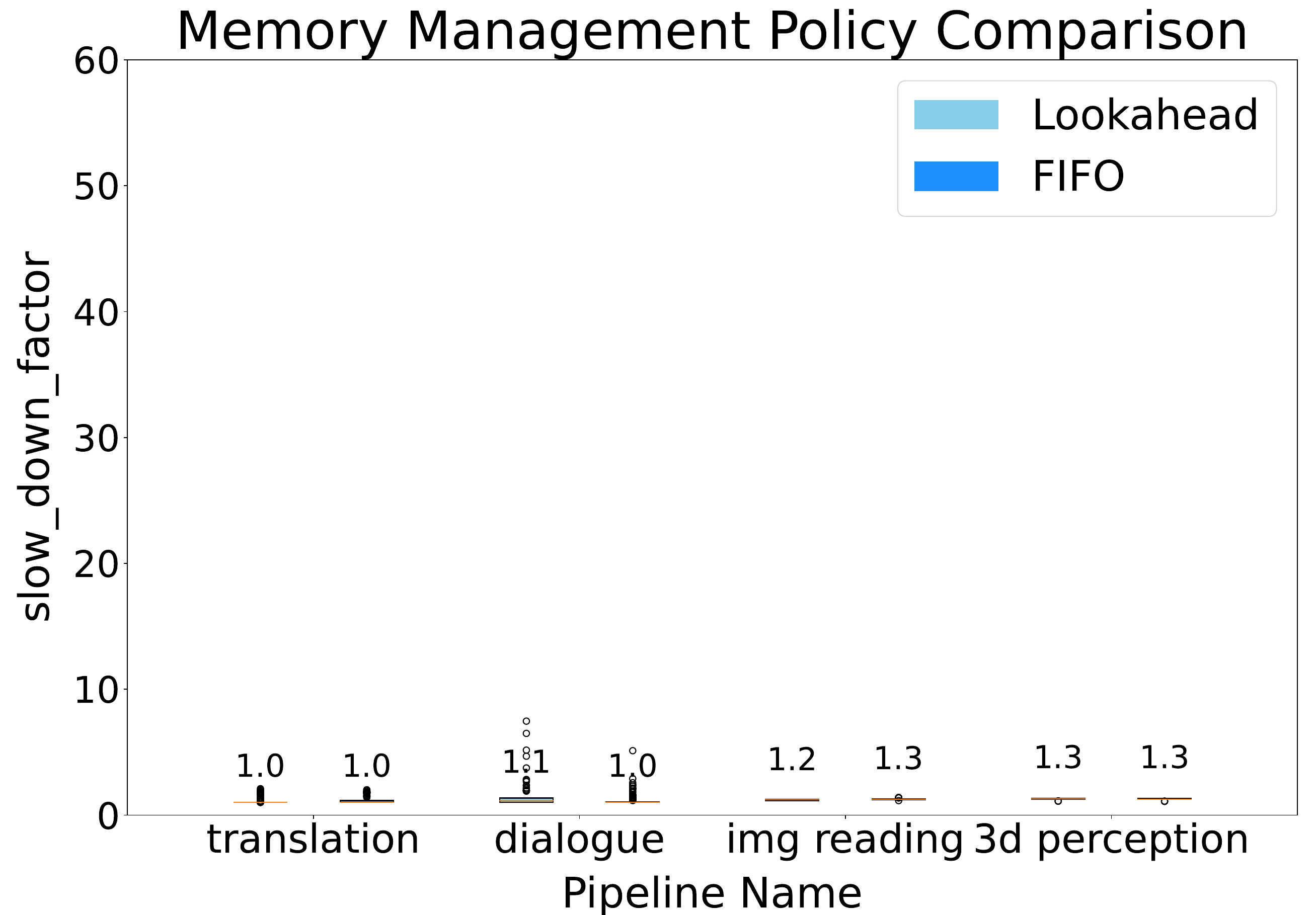}
         \caption{GPU Memory Management policies comparison under low request rate}
         \label{fig:eviction_low_rate}
     \end{subfigure}
     \hfill
     \begin{subfigure}[b]{0.23\textwidth}
         \centering
         \includegraphics[width=\textwidth]{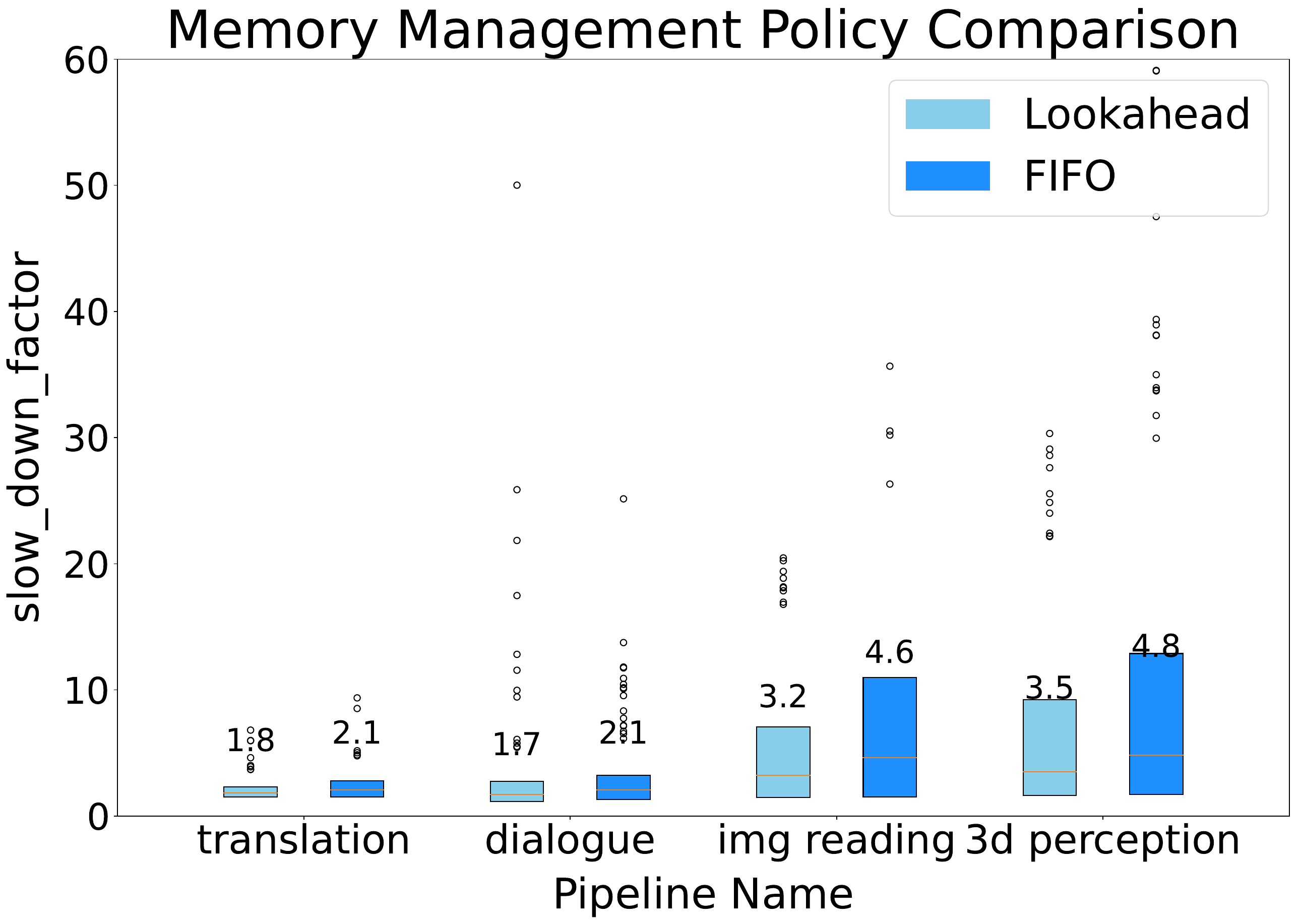}
         \caption{GPU Memory Management policies comparison under high request rate}
         \label{fig:eviction_high_rate}
     \end{subfigure}
\caption{GPU Memory Management policies comparison}
\label{fig:memory_policy_comparison}
\end{figure}

}

%% file: 7_related_work.tex
\section{Related work}
The study of scheduling and cache management for ML has been addressed in several prior works. Commercial systems such as TensorFlow Serving\cite{tensorFlowServing}, NVIDIA Triton\cite{triton}, TFX Platform \cite{tfx} support ML application deployment in production environments optimized for performance.   

Prior work covers a range of perspectives.  ML model optimization is the focus in ~\cite{Nimble, Orca} whereas hardware management is the more central emphasis in ~\cite{clipper, Clockwork, alpaserve, gpulet, singularity, antman, walle, mlaas_in_the_wild}. Nimble~\cite{Nimble} compiles and optimizes dynamic deep learning models to portable VM-runtime. Orca~\cite{Orca} performs iteration-level optimization to the transformer model to improve latency and throughput. Clockwork~\cite{Clockwork} employs a fine-grained approach to manage the model in memory and during execution, and minimize tail latency to achieve predictable performance. To optimize for model serving SLOs,  AlpaServe~\cite{alpaserve} utilizes model parallelism and statistical multiplexing, while the gpulet~\cite{gpulet} scheduler takes the approach of spatio-temporal sharing of partitioned hardware resources to support heterogeneous ML models. Singularity~\cite{singularity}, Antman~\cite{antman}, Walle~\cite{walle} and PAI~\cite{mlaas_in_the_wild} schedule ML tasks in large-scale cloud production platforms. Cloud scalability encourages this type of elastic upscaling and downscaling, but these methods cannot directly transfer to small edge cluster with ML tasks that have large models and employ GPU accelerators. Additionally, this work focuses primarily on workflows with just a single ML task per request, and generalization to  \staticgraph  workflows would be a significant undertaking.

Scheduling for workflows with inter-job dependencies has been explored in many prior efforts ~\cite{unearth_inter_job_dependency, Apollo, HEFT, batchy}. Wing~\cite{unearth_inter_job_dependency} focuses on data-analytic queries; Apollo~\cite{Apollo} focuses on data-processing jobs with dependencies in production cluster; HEFT~\cite{HEFT} schedules for general-purpose computation workflow on heterogeneous processors. 
Batchy~\cite{batchy} is a general-purpose dataflow graph scheduler that uses batch-processing profiles to balance the efficiency and latency under SLO constraints. The estimation-based job planning and deferred correction mechanism used in Apollo~\cite{Apollo} resembles the scheduling scheme in Compass; while the task assignment algorithm in HEFT~\cite{HEFT} resembles Compass's job planning algorithm. However, direct application of these methods to the ML workflows in Compass does not give latency-optimal scheduling result because they do not consider ML model re-use, GPU memory consumption and collocation of ML models. 
 
Compass's graphical workflows are similar to those considered in ~\cite{video_storm, nexus, elasticStream, InferLine, orion}. Video storm\cite{video_storm} and Nexus \cite{nexus} schedule pipeline-structure workflows for video stream. Orion~\cite{orion} schedules serverless DAGs with right-sizing, bundling and pre-warming techniques to meet latency SLOs. Stream Processing Engines (SPEs) scheduler~\cite{elasticStream} is similar with Orion~\cite{orion} in that both use workload prediction model to estimate the worker waittime and provision the workers accordingly. These provisioning methods rely on workload prediction, but potentially sacrifice latency when the workload distribution drifts or if a burst of jobs puts the system under resource stress. InferLine\cite{InferLine} introduces a low-frequency combinatorial planner and a high-frequency auto-scaling tuner mechanism that it uses to upscale and downscale the number of workers as load varies. The process of worker selection for provision and up-scaling is less addressed in InferLine~\cite{InferLine}, which becomes nontrivial when the model sizes are significantly larger and when there are only limited GPU memories. Compass addresses this scenario by scheduler and resource management co-design, which optimizes the task assignment while balancing the co-location of ML model objects in Compass-managed GPU memory.

%% file: 8_conclusion.tex
\section{Conclusion}
Compass manages memory and schedules tasks in support of latency-sensitive ML workflows structured as DFGs. A novel  scheduling-aware eviction policy yields high GPU cache hit rates, avoiding unnecessary model fetching and eviction. Our evaluation compared Compass with state of the art schedulers under a variety of request rates and workload patterns, showing  2x to 6x speedup with similar resource use, while leaving excess servers completely idle.
Compass also turns out to perform well with respect to energy consumption.

%% file: 10_Acknowledgments.tex
\section{Acknowledgments}

We are grateful to Professor Luís Rodrigues for his many insightful comments and to Microsoft Corporation, Siemens Corporation and the Cornell Institute for Digital Agriculture for funding the effort.